\documentclass[11pt, a4paper]{article}
\usepackage[utf8]{inputenc}

\usepackage[margin=1in]{geometry} 
\usepackage[auth-sc,affil-sl]{authblk}
\usepackage{cite}

\usepackage{amsmath}
\usepackage{amsthm}
\usepackage{amssymb}
\usepackage{bbm}

\usepackage{graphicx}
\usepackage{caption}
\usepackage{subcaption}
\usepackage{float}

\usepackage{hyperref}

\usepackage{appendix}
\usepackage{booktabs}

\newcommand{\be}{\begin{equation}}
\newcommand{\ee}{\end{equation}}
\newcommand{\bal}{\begin{aligned}}
\newcommand{\eal}{\end{aligned}}
\newcommand{\mathN}{\mathcal{N}}

\DeclareMathOperator*{\argmin}{arg\,min}

\usepackage{color}
\definecolor{colorML}{rgb}{0, .5, 1}
\definecolor{colorGG}{rgb}{.7,.5,.9}

\title{\Huge Learning new physics efficiently with nonparametric methods}
\author[1,2]{Marco Letizia}
\author[1]{Gianvito Losapio}
\author[1]{Marco Rando}
\author[3,4,5]{Gaia Grosso}
\author[3]{Andrea Wulzer}
\author[5]{Maurizio Pierini}
\author[3,4]{Marco Zanetti}
\author[1,6,7]{Lorenzo Rosasco}
\affil[1]{MaLGa - DIBRIS,  Universit\`{a} di Genova, Genova, Italy}
\affil[2]{INFN,  Sezione di Genova, Genova, Italy}
\affil[3]{Dipartimento di Fisica e Astronomia,  Universit\`{a} di Padova, Padova, Italy}
\affil[4]{INFN,  Sezione di Padova, Padova, Italy}
\affil[5]{Experimental Physics Department, CERN, Geneva, Switzerland}
\affil[6]{CBMM,  Massachusetts Institute of Technology, Cambridge, MA, USA}
\affil[7]{Istituto Italiano di Tecnologia, Genova, Italy}

\date{}

\begin{document}

\maketitle

\abstract{We present a  machine learning approach for model-independent new physics searches. The corresponding algorithm  is powered by recent large-scale implementations of kernel methods, nonparametric learning algorithms that can approximate any continuous function given enough data.  Based on the original proposal by D'Agnolo and Wulzer \cite{DAgnolo:2018cun},  the model evaluates the compatibility between experimental data and a reference model, by implementing a hypothesis testing procedure based on the likelihood ratio.  Model-independence is enforced by avoiding any prior assumption about the presence or shape of new physics components in the measurements.  We show that our approach has dramatic advantages compared to neural network implementations in terms of training times and computational resources, while maintaining comparable performances.  In particular,  we conduct our tests on higher dimensional datasets,  a step forward with respect to previous studies.}

%%%%%%%%

\section{Introduction}\label{sec:intro}

Experimental observations and convincing conceptual arguments indicate that the present understanding of fundamental physics is not complete. 
Our theoretical formulation of the fundamental laws of Nature, the Standard Model, has been predicting with extremely high precision an impressive amount of data collected at past and ongoing experiments. On the other hand, the Standard Model does not provide answer to a multitude of questions including the origin of the electroweak scale,  the mass of neutrinos, the flavour structure in the quark, lepton and neutrino sectors, and is unable to account for observed  phenomena like the origin and the composition of the dark matter of the baryon asymmetry in the Universe. Further, it does not provide a microscopic description of gravity. These considerations guarantee the existence of more fundamental laws of Nature waiting to be unveiled. In order to access these laws, we must search the experimental data for phenomena that depart from the Standard Model predictions.

Currently, the most common searching strategy is to test the data for the presence of specific new physics models, one at the time.
Each search is then optimized to be sensitive to the features specific of the considered new physics scenario.
This approach is in general insensitive to sources of discrepancy that differ from those considered.
There is therefore a strong effort in developing analysis strategies that are agnostic about the nature of potential new physics and thus complementary to the model-dependent approaches described above
\cite{Choudalakis:2011qn,D0:2000vuh,D0:2000dnz,H1:2004rlm,H1:2008aak,Asadi:2017qon,CDF:2007iou,CDF:2008voc,CMS:2008gya,CMS:2017yoc,ATLAS:2014sxa,ATLAS:2017irs}.  Ideally,  this type of analysis should be sensitive to generic departures from a given reference model.  In practice, this is a challenge given the complexity of the experimental data in modern experiments and the fact that the new physics signal is expected to be ``small" and/or located in a region of the input features which is already populated by events predicted by the reference model.
Recently, there has been a strong push towards developing solutions based on machine learning for (partial or full) model-independent searches in high energy physics \cite{Weisser:2016cnc,Cerri:2018anq,DAgnolo:2018cun,DAgnolo:2019vbw,DeSimone:2018efk,Farina:2018fyg,Collins:2018epr,Blance:2019ibf,Hajer:2018kqm,Heimel:2018mkt,Collins:2019jip,Nachman:2020lpy,Andreassen:2020nkr,Amram:2020ykb,Dillon:2020quc,Cheng:2020dal,Khosa:2020qrz,Nachman:2020ccu,Park:2020pak,Bortolato:2021zic,Finke:2021sdf,Gonski:2021jek,Hallin:2021wme,Ostdiek:2021bem,Chakravarti:2021svb}.  

In this work we present a novel machine learning implementation of the analysis strategy proposed by D'Agnolo et al. \cite{DAgnolo:2018cun,DAgnolo:2019vbw}. 
The aim of this strategy is to compute the log-likelihood-ratio test statistics without specifying the alternative new physics hypothesis a priori. Towards this end, a neural network model was used in \cite{DAgnolo:2018cun,DAgnolo:2019vbw}  to learn the alternative hypothesis directly from the data while the log-likelihood-ratio was maximized to get an optimal test statistics.
The strategy assumes that a sample of events representing the Standard Model hypothesis (``reference" sample) is available and that its size is much larger than the one of the experimental data, so that the only relevant statistical uncertainties are those of the data themselves.
In the new implementation presented here,  neural networks are replaced by kernel methods.  Kernel methods are nonparametric algorithms that can approximate any continuous function given enough data.  Recent large-scale implementations \cite{falkonlibrary2020} provides fast and efficient solvers even with very large data-sets.
This is relevant since a key bottleneck of the neural network model used in Ref.~\cite{DAgnolo:2018cun,DAgnolo:2019vbw} is the extremely long training time,  even on low dimensional problems.
The solution we propose solves this issue by delivering comparable performances with orders of magnitude gain in training times,  see Table \ref{table:tr_times}. We demonstrate the viability of the  framework by testing on particle physics datasets of increasing dimensionality,  a further step forward with respect to previous studies.

We note that the ideas recently proposed in Ref.~\cite{Chakravarti:2021svb} share some similarities to our approach.  Indeed, the authors of Ref.~\cite{Chakravarti:2021svb} developed a model-independent strategy based on classifiers to perform hypothesis testing on Standard Model samples and experimental measurements.  However,  they implement a train-test split of the data for the reconstruction of the test statistics and for inference.  This is a major difference with respect to our approach, where the distribution employed for the evaluation of the test statistics is the one that best fits the very same set of data on which the test has to be performed,  in accordance with the maximum likelihood philosophy. 
Moreover,  while their approach permits to estimate the distribution of the test statistics with a single training of a classifier,  only half of the experimental data is used for new physics detection.  
A in-depth comparison of the two models will be explored in future works.

The rest of the paper is organized as follows.  In Section \ref{sec:framework} we introduce the main statistical framework at the basis of this work,  elaborating on the discussion in Ref.~\cite{DAgnolo:2018cun}.  In Section \ref{sec:model}, we discuss the different aspects of the proposed model,  in particular the underlying machine learning algorithm.  In Section \ref{sec:exp}, we test the algorithm on realistic simulated datasets in various dimensions and we explicitly compare our proposal with the neural network models in Ref.~\cite{DAgnolo:2018cun,DAgnolo:2019vbw}.  Finally in Section \ref{sec:conclusions}, we lay out our conclusions and discuss future developments.  In the appendices, we review some background material and present other complementary experiments.

%%%%%%%%%

\section{Statistical foundations}\label{sec:framework}

In this section, we reprise and elaborate the main ideas in Refs.~\cite{DAgnolo:2018cun,DAgnolo:2019vbw}, tackling the problem of testing the data for the presence of new physics with tools from statistics and machine learning. 

We start by assuming that an experiment is performed and its outcome can be described by a multivariate random variable $x$. A physical model corresponds to an ensemble of mathematical laws characterizing a distribution for $x$. In this view, we denote by $p(x|0)$ the distribution of the measurements as described by the Standard Model and by $p(x|1)$ the unknown true distribution of the data.  Discovering new physics will be cast as the problem of {\em testing} whether the latter coincides with the former or not.  

The  distribution $p(x|0)$ is essentially known.  
Although not analytically computable in most  high energy physics applications, it can be sampled via Monte Carlo simulations or extracted using control regions with data driven techniques. In the following, we denote one such set of independent and identically distributed random variables ($i.i.d.$) by 
\be
S_0=\{x_i\}_{i=1}^{\mathN_0},\;\textrm{with}\; x_i\overset{i.i.d.}{\sim} p(x|0),
\ee 
and  the actual measured data by,
\be
S_1=\{x_i\}_{i=1}^{\mathN_1},\;\textrm{with}\; x_i\overset{i.i.d.}{\sim} p(x|1).
\ee 
It should be pointed out that in real applications one would also consider the uncertainties affecting the knowledge of the reference model. Similarly to Refs.~\cite{DAgnolo:2018cun,DAgnolo:2019vbw}, we will  assume ${\cal{N}}_0 \gg {\cal{N}}_1$ so that the statistical uncertainties on the reference sample can be neglected. 
It should be possible to include systematic uncertainties as nuisance parameters,  as shown in Ref.~\cite{dAgnolo:2021aun} for the neural network implementation.  However, we assume that the systematic uncertainties are negligible in what follows and leave this aspect to future works. 

The idea in Ref~\cite{DAgnolo:2018cun} is to translate the maximization of the log-likelihood-ratio test into a machine learning problem, where the null hypothesis characterising one of the likelihood terms is the reference hypothesis (namely the Standard Model) and the alternative hypothesis characterising the other likelihood term is unspecified a priori and learnt from the data themselves during the training. The test statistic obtained in this way is therefore a good approximation of the optimal test statistic according to the Neyman-Pearson lemma.\\
We define the likelihood of the data $S_1$ under a generic hypothesis $H$ as
\be\label{extlikelihood}
\bal
\mathcal{L}(S_1,H)&=\frac{e^{-N(H)} N(H)^{\mathN_1}}{\mathN_1!}\prod_{x=1}^{\mathN_1} p(x|H)\\
&=\frac{e^{-N(H)}}{\mathN_1!}\prod_{x}n(x|H), 
\eal
\ee
where 
\be\label{diffdistr}
n(x|H)=N(H) p(x|H)
\ee
is the data distribution normalized to the expected number of events  
\be
N(H)=\int n(x|H)\,dx.
\ee
As already said, $p(x|0)$ is essentially known and well represented by the reference sample while $p(x|1)$ is not and thus its exact form must be replaced by a family of distributions $p_w(x|1)$, parametrized by a set of trainable variables $w$. 
We can write the  likelihood ratio test statistics as,
\be\label{test_st}
\bal
t_w(S_1)&=-2\log \frac{\mathcal{L}_w(S_1,0)}{\mathcal{L}(S_1,1)}\\
&=-2\log \left[e^{N_{w}(1)-N(0)} \prod_{x=1}^{\mathN_1}\frac{n(x|0)}{n_w(x|1)}\right]\\
&=-2\left[N_w(1)-N(0)-\sum_{x=1}^{\mathN_1} \log\frac{n_w(x|1)}{n(x|0)} \right].
\eal
\ee
and optmize it by maximizing over the set of parameters $w$. 
The original proposal in Ref.~\cite{DAgnolo:2018cun} suggested to exploit the ability of neural networks as universal approximators to define a family of functions describing the log-ratio of the density distributions in Eq.~\eqref{test_st}
\be
f_w(x) = \log\frac{n_w(x|1)}{n(x|0)}.
\ee
As discussed below the same approach can be taken replacing neural networks with other machine learning approaches, e.g. kernel methods. 
Following the above reasoning, the maximum of the test statistic could then be rewritten as the minimum of a loss function $L(S_1, f_w)$
\be\label{loss_nplm}
\bal
t_{\hat w}(S_1) &= \max_{w}\,t_{w}(S_1)\\
			&=-2 \min_{w}\, L(S_1, f_w)\\
			&=-2\min_{w}\, \left[ \sum_{S_0}\frac{N(0)}{{\cal N}_0}(e^{f_w(x)}-1) - \sum_{S_1} f_w(x) \right]
\eal
\ee
and the set of parameters $\hat w$ which maximizes $t_w(S_1)$ 
\be\label{nratio}
n_{\widehat w}(x|1)=n(x|0)e^{f_{\hat w}(x)}\approx n(x|1)
\ee
provides also the best approximation of the true underlying data distribution and with it a first insight on the source and shape of the discrepancy, if present. Note that the loss defined by Eq.~\eqref{loss_nplm} is unbounded from below.  In Ref.~\cite{DAgnolo:2018cun} a regularization parameter is introduced as a hard upper bound (weight clipping) on the magnitude of the parameters $w$.

\subsection{Designing a classifier for hypothesis testing}
In this work we develop the above ideas  considering a different loss function,  namely a weighted cross-entropy (logistic) loss function. This was a possibility mentioned as a viable alternative in Ref.~\cite{DAgnolo:2018cun} that   we indeed show to  yields several advantages. 
To estimate the ratio in Eq.\eqref{nratio} we train a binary classifier on $S=S_0\cup S_1$ using a weighted cross-entropy loss
\be\label{wbce}
\ell(y,f(x))=a_0 (1-y) \log \left(1+e^{f(x)}\right)+a_1 y \log \left(1+e^{-f(x)}\right).
\ee
where $y$ is the class label and takes value zero for $S_0$ and one for $S_1$.
The classifier is obtained minimizing an empirical criterion
\be\label{ERM}
\hat{L}(f_w)=\frac{1}{\mathN} \sum_{i=1}^\mathN \ell(y,f_w(x)),
\ee
over a suitable class of machine learning models $f_w$. If such a  models class is sufficiently rich,  in the large sample limit we would recover a minimizer of the expected risk
\be\label{expRisk}
L(f)=\int \ell (y,f(x)) dp(x,y),
\ee
where $p(x,y)$ is the joint data distribution. 
By a standard computation (see Appendix \ref{app:SLT}), the function minimizing the expected risk in Eq.~\eqref{expRisk} can be shown to be
\be\label{pratio}
f^*(x)=\log\left(\frac{p(1|x)}{p(0|x)}\frac{a_1}{a_0}\right),
\ee
that,  by Bayes theorem and Eq.\eqref{diffdistr}, we can rewrite as
\be
f^*(x)=\log\left(\frac{p(x|1)}{p(x|0)}\frac{p(1)}{p(0)}\frac{a_1}{a_0}\right)=\log\left(\frac{n(x|1)}{n(x|0)}\frac{N(0)}{N(1)}\frac{p(1)}{p(0)}\frac{a_1}{a_0}\right).
\ee
From the above expression and choosing the weights so that
\be
\frac{a_1}{a_0}=\frac{N(1)}{N(0)}\frac{p(0)}{p(1)},
\ee
Eq.\eqref{pratio} reduces to Eq.\eqref{nratio}, as desired.  In practice,  the above condition can be satisfied only approximately, since it depends on quantities we do not know. Hence, we first estimate the class priors using the empirical class frequencies, $p(y)\approx \mathN_y/\mathN$ with $\mathN=\mathN_0+\mathN_1$ and obtain
\be
\frac{a_1}{a_0}\approx\frac{\hat{a}_1}{\hat{a}_0}=\frac{N(1)}{N(0)}\frac{\mathN_0}{\mathN_1}.
\ee
Then, we approximate the number of expected events in the alternative hypothesis with the actual number of experimental measurements $N(1)\approx\mathN_1$.\footnote{This is exact on average, since $\mathN_1\sim \text{Pois}(N(1))$.}
The following expression of the weights can then be used in practice,
\be\label{coeff}
\frac{\hat{a}_1}{\hat{a}_0}=\frac{\mathN_0}{N(0)}. 
\ee
To reconstruct the test statistics in Eq.\eqref{test_st}, the number of expected events in the alternative hypothesis needs to be computed.  Using the density ratio in Eq.\eqref{nratio}, we have that 
\be
\bal
&N_w(1)=\int n_w (x|1) \, dx= \int n(x|0) \, e^{f_w (x)}  dx,\\
&\textrm{with}\quad \frac{n_w(x|1)}{n(x|0)}=e^{f_w(x)}.
\eal
\ee
Since the reference distribution $n(x|0)$ is not known analytically, we can estimate the above expression  using a Monte Carlo approximation considering
\be\label{recoN1}
N_w(1)\approx \frac{N(0)}{\mathN_0} \sum_{x\in S_0} e^{f_w(x)}.
\ee
Using Eq.~\eqref{recoN1}, the test statistics in Eq.\eqref{test_st} can be written as
\be\label{t}
t_{\hat{w}}(S_1)=-2\left[\frac{N(0)}{\mathN_0}\sum_{x\in S_0}\left(1-e^{f_{\hat{w}}(x)}\right)    +\sum_{x\in S_1} f_{\hat{w}}(x) \right]
\ee
recovering the original result from Ref.~\cite{DAgnolo:2018cun}.

The main conceptual difference with respect to the original solution in Ref.~\cite{DAgnolo:2018cun} lays on the computation of the test statistic.  When using the loss in Eq.~\eqref{loss_nplm} the test statistic can be directly obtained from the value of the loss function at the end of the training. When using the cross-entropy loss, each term of the log-likelihood-ratio test is  calculated separately and then combined, see Eq.~\eqref{t}. This could be a problem if the optimality of the minimization procedure is not ensured. More precisely, in the first case the minimum found at the end of the training is by construction the one maximizing the log-likelihood-ratio test, while this is guaranteed only in the asymptotic limit in the second case. On the other hand, as noted before, the loss function in Eq.~\eqref{loss_nplm} is less well behaved from a mathematical standpoint, making optimization during training less trivial. Interestingly,  both loss functions are designed to estimate the same density ration,  and  in practice  we show that they obtain comparable performances   in terms of sensitivity to new physics.

We conclude noting that the value of the  test statistic $ t_{\widehat w}(S_1)$ is a random variable itself following a distribution $p(t|H)$. The level of significance associated to a value of the test statistic is computed as a $p$-value of the test statistic with respect to its distribution under the null hypothesis
\be\label{pvalue}
p_{S_1} = \int_{t(S_1)}^\infty p(t|0)\,dt.
\ee
This can be further rewritten as a Z-score
\be\label{Zobs}
Z_{obs}(S_1)=\Phi^{-1}(1-p_{S_1}),
\ee
where $\Phi^{-1}$ is the quantile of a Normal distribution.  In this way $Z_{obs}$ is expressed in units of standard deviations.
Following Ref.~\cite{DAgnolo:2018cun}, by leveraging the possibility to sample from the reference distribution, we choose to reconstruct $p(t|0)$ by estimating the likelihood ratio test statistics on a number $N_{toy}$ of toy experiments run on pseudo datasets extracted from the reference sample. The latters have the same statistics of the actual data but do not have any genuine new physics component.

\paragraph{Class imbalance.}
To accurately represent the reference distribution, it is preferable to consider a large reference sample,  while the number of experimental samples is determined by the parameters of the experiment, specifically its luminosity.  This leads to an imbalanced classification problem and a natural approach is to re-weight the  loss using the inverse class frequencies $\mathN_y$.  The true number of expected events differ from the number of events in the reference hypothesis by the number of expected new physics events, i.e., $N(1)=N(0)+N(S)$. Then, one has that $\mathN_1\sim \text{Pois}(N(0)+N(S))$.  From both the experimental and theoretical points of view,  it is reasonable to assume that $N(S)\ll N(0)$.  Therefore,  one has that $\mathN_1\approx N(0)$.  Hence,  by using the weight in Eq.\eqref{coeff}, besides recovering the desired target function,  we solve potential issues related with an imbalanced dataset, while keeping the statistical advantage of having a large reference sample. 

\subsection{Analysis strategy}\label{training_scheme}
The complete analysis strategy can be summarized in three steps:
\begin{itemize}
\item the test statistic distribution is empirically built by running the training on $N_{toy}=\mathcal{O}(100)$ toy experiments for which both the training sample $S_1$ and $S_0$ are generated according to the null hypothesis.
\item One last training is performed on the dataset of interest $S_1$ for which the true underlying hypothesis is unknown and the test statistic value $t(S_1)$ is evaluated.
\item The $p$-value corresponding to $t(S_1)$ is computed with respect to the test statistic distribution under the null hypothesis, studied at step 1.
\end{itemize}
If a statistically significant deviation from the reference data is found,  the nature of the discrepancy can be further characterized by inspecting the learned density ratio in Eq.\eqref{nratio}.  This quantity is expected to be approximately zero if no disagreement is found and it can be inspected as a function of the input features or their combinations.\\

\paragraph{Asymptotic formula}\label{par:asympt}
Typically, for an accurate estimation of $p(t|0)$, the empirical distribution of the test statistic under the reference hypothesis has to be reconstructed using a large number of toy experiments and this might be practically unfeasible.  
If the value of $t(S_1)$ falls outside of the range of the empirical distribution the p-value cannot be computed and only a lower bound can be set.  
Inspired by the results by Wald and Wilks \cite{Wilks:1938dza,wald1943tests,Cowan:2010js} characterizing the asymptotic behavior of the log-likelihood test statistics,  we approximate the null distribution with a $\chi^2$ distribution.  We use the toy-based empirical estimate to determine the degrees of freedom of the $\chi^2$ distribution and we test the compatibility of the empirical test statistic distribution with the $\chi^2$ hypothesis using a Kolmogorov-Smirnov test. 
This approximation holds well in almost all instances of our model. We did not explore this aspect in details but we present a counterexample towards the end of Section \ref{sec:exp}.  The same approximation is also used in the neural network model of \cite{DAgnolo:2018cun,DAgnolo:2019vbw}.  It is worth specifying that, in real-life scenarios,  if the  p-value computed in this way would imply a discovery,  one would run additional toys to obtain an accurate empirical estimation by brute-force exploitation of the large-scale computing resources typically accessible by the LHC collaborations.

\section{Scalable nonparametric learning with  Kernels} \label{sec:model}
As mentioned before, a rich model class is needed to effectively detect new physics clues in the data.
In this work, we consider kernel methods\cite{hastie01statisticallearning,falkonlibrary2020} of the form
\be\label{kernel_mod}
f_w(x)=\sum_{i=1}^\mathN w_i k_\gamma(x,x_i).
\ee
Here $k_\gamma (x,x_i)$ is the kernel function and $\gamma$  some hyper-parameter. In our experiments, we consider the Gaussian kernel
\be
k_{\sigma}(x,x')=e^{-\Vert x-x'\Vert^2/2\sigma^2},
\ee
so that  $f_w$ corresponds to  a linear combination of Gaussians of prescribed width $\gamma$, centered at the input points.  Such an approach is called nonparametric because the number of parameters corresponds to the number of data points: the more the data,  the more the parameters.  Indeed, this makes  kernel methods universal in the large sample limit,  in the sense that they can recover any continuous function \cite{JMLR:v7:micchelli06a,christmann2008support}. 
The computational complexity to determine  a function as  in Eq.~\eqref{kernel_mod} is typically cubic 
in time and quadratic in space with respect to the number of points. 
These costs prevent the application of basic  solvers in large-scale setting,  and some approximation is needed. 
Towards this end we consider Falkon \cite{falkonlibrary2020}, which replaces Eq.~\eqref{kernel_mod}  by 
\be\label{kernel_sol}
f_w(x)=\sum_{i=1}^M w_i k_{\sigma}(x,\tilde x_i),
\ee
where $\{\tilde{x}_1,..., \tilde{x}_M\} \subset \{x_1,...,x_\mathN\}$ are called Nystr\"{o}m centers and are sampled uniformly at random from the input data, with $M$ an hyper-parameter to be chosem.  Notably, the corresponding solution can be shown to be with high probability as accurate as the original exact one while computable with only a small fraction of computational resources \cite{sun2018but,rudi2021generalization,bach2013sharp,rudi2016more,calandriello2019statistical,li2019towards}. We defer further details to the appendices. 

\paragraph{Algorithm training}
The model's weights in Eq.~\eqref{kernel_sol} are computed to minimize the empirical error~\eqref{ERM} defined by the weighted cross-entropy loss introduced before.
Since, the kernel model can be very rich, the search of the best model is done considering
\be\label{reg_ERM}
\hat  L(f_w)+\lambda R(f_w),
\ee
where the first term is the empirical risk, while $R(f)$ is a regularization term
\be
R(f_w)=\sum_{ij} w_i w_j k_{\sigma}(x_i,x_j).
\ee
constraining the complexity of the model \cite{shalev2014understanding}. Problem~\eqref{reg_ERM} is then solved  by an approximate Newton iteration \cite{falkonlibrary2020}. 

\paragraph{Hyper-parameters tuning}
The number of Nystr\"{o}m centers ($M$), the bandwidth of the Gaussian kernel $\sigma$ and the regularization parameter $\lambda$ are the main hyper-parameters of the model.
The number of centers $M$ determines the number of Gaussians,  hence it has an impact on the accuracy and on the computational cost; studies suggests that optimal statistical bounds can be achieved already with $M=\mathcal{O}(\sqrt{\mathN})$ \cite{rudi2016more,marteauferey2019globally}.  
On the other hand, by varying the hyper-parameters $\sigma$ and $\lambda$, more or less complex functions can be selected.  
For large $\lambda$ or $\sigma$ the model simplifies and tends to be linear, while for small values it tends to fit the statistical fluctuations in the data.  

The values of $M$, $\sigma$ and $\lambda$ affect the distribution of the test statistic under the reference hypothesis. In particular we observe that the test statistic distribution obtained with different choices of the hyper-parameters always fits a $\chi^2$ distribution with a number of degrees of freedom determined empirically as explained in Section \ref{par:asympt}.
 More complex functions cause the distribution of the test statistic to move to higher values (see Figure~\ref{fig:univ_tuning}).

On the $M$ direction,  a stable configuration is eventually reached and this information can be used to select a proper trade-off value for $M$ (see for instance Figure~\ref{fig:univ_tuning_all}); conversely there is not clear indication on how to choose the values of $\sigma$ and $\lambda$. 
The bandwidth $\sigma$ is related to the resolution of the model and its ability to fit statistical fluctuations in the data. To estimate the relevant scales of the problem and find a good trade-off between complexity and smoothness, we look at the distribution of the pairwise (Euclidean) distance in the reference data. 
We then fix $\sigma$ approximately as the 90th percentile (see Appendix \ref{app:univ} and Figure \ref{fig:varsigma} for further details). 
Finally, $\lambda$ determines the weight of the penalty term in the loss function, which constraint the magnitude of the trainable weights, and avoid instabilities during the training. We take $\lambda$ as small as possible so that the impact on the weight magnitude is minimum, while maintaining the algorithm numerically stable.  

Summarizing, the hyper-parameter tuning protocol is composed by the following three steps:
\begin{itemize}
\item We consider a number of centers greater or equal to $\sqrt{\mathN}$,  
with the criteria that more centers could improve accuracy but at the cost of losing efficiency. 
\item We then fix $\sigma$ approximately as its 90th percentile of the pairwise distance distribution. 
\item We take $\lambda$ as small as possible while maintaining a numerically stable algorithm.
\end{itemize}

Similarly to the tuning procedure introduced in Ref.~\cite{DAgnolo:2019vbw} for the neural networks, the outlined directives for hyper-parameters selection rely on the reference data alone, preserving model-independence. 

We tested this heuristic performing several experiments on the toy scenario presented in Appendix \ref{app:univ}.  In particular,  we verified that it gives rise to instances that demonstrate good performances, in terms of sensitivity to new physics clues,  across different types of signal.  We also verified that the results are robust against small variations of the chosen hyper-parameters.  When applied to the final experiments presented in the following section,  we followed the prescription given above without any fine tuning that might introduce a bias that favors the specific dataset considered. 

\paragraph{Assessing the algorithm performances}
Following Ref.~\cite{DAgnolo:2018cun}, in order to evaluate different models on benchmark cases it is useful to introduce the ideal significance $Z_{id}$.,  i.e.,  the value of the median Z-score that 
is obtained by using the exact (ideal) likelihood ratio test statistics:
\be
t_{id}(\cdot)=-2\log\frac{\mathcal{L}(\cdot,1)}{\mathcal{L}(\cdot,0)}.
\ee
Typically,  this quantity cannot be computed exactly since the likelihoods are not known analytically. 
We can however obtain an accurate estimate $\hat{Z}_{id}$ using simulated data and model-dependent analyses that leverage what is known about the type of new physics in the data. 
We will report how $\hat{Z}_{id}$ has been computed for every experiment.

%%%%%%%%%

\section{Experiments}\label{sec:exp}
In this section,  we apply the proposed approach to three realistic simulated high energy physics datasets with an increasing number of dimensions.  Each dataset is made of two classes: a reference class,  containing events following the Standard Model,  and a data class,  made of reference events with the injection of a new physics signal.  Each case includes a set of features given by kinematical variables as measured by the particle detectors (plus additional quantities when available, such as reconstructed missing momenta and b-tagging information) that we call \textit{low-level features}.  From the knowledge of the intermediate physics processes, one can compute additional \textit{high-level features} that are functions of low-level ones and posses a higher discriminative power. \footnote{We borrow this nomenclature from Ref.~\cite{Baldi:2014kfa}.} The different features are used to test the flexibility of the model.  
The pipeline for training and tuning our method is that described in Section \ref{sec:model}.

\subsection{Datasets}
Here,  we briefly review some properties of the datasets,  how $\hat{Z}_{id}$ is computed and the parameters chosen for the experiments.  We refer the reader to Ref.~\cite{DAgnolo:2019vbw,Baldi:2014kfa} for further details.

\paragraph{DIMUON}
This is a five dimensional simulated dataset that was introduced in Ref.~\cite{DAgnolo:2019vbw} and it is composed of simulated LHC collision events producing two muons in the final state $pp\rightarrow\mu^+\mu^-$,  at a center-of-mass energy of 13 TeV.  The low-level features are the transverse momenta and pseudorapidities of the two muons and their relative azimuthal angle, i.e., $x=[p_{T1},p_{T2},\eta_1,\eta_2,\Delta\phi]$.   We consider two types of new physics contributions: the first one is a new vector boson ($Z'$) for which we study different mass values ($m_{Z'}=200,300$ and 600 GeV); the second one is instead a non-resonant signal obtained by adding a four fermions contact interaction to the Standard Model lagrangian\\
\be
\frac{c_W}{\Lambda}J_{L\mu}^aJ_{La}^{\mu}
\ee
where $J_{La}^{\mu}$ is the $\rm{SU(2)}_L$ Standard Model current, the energy scale $\Lambda$ is fixed at $1$ TeV and the Wilson coefficient $c_W$ determining the coupling strength can be chosen between three values ($c_W=1, 1.2$ and 1.5 $\textrm{TeV}^{-2}$). 
For both types of signal the invariant mass of the two muons is the most discriminant non trivial combination of the kinematic variables describing the system so we consider it as a high-level feature.
We fix $N(0)=2\times 10^4$ expected events in the reference hypothesis and the size of the reference sample is $\mathN_0=10^5$, unless specified otherwise.  We vary the number of expected signal events in the range $N(S)\in[6,80]$.  We selected the following hyper-parameters: $(M,\sigma,\lambda)=(2\times 10^4,3,10^{-6})$.

The ideal significance is estimated via a cut-and-count strategy in the invariant mass $m_{\ell\ell}$ distribution around $m_{Z'}$ for the resonant signal, while a likelihood ratio test on the binned $m_{\ell\ell}$ distribution is used for the non-resonant case. 

\paragraph{SUSY}
The SUSY dataset  \cite{Baldi:2014kfa} is composed of simulated LHC collision events in which the final state is made of two charged leptons $\ell\ell$ and missing momentum. 
The latter is given, in the Standard Model, by two neutrinos coming from the fully leptonic decay of the two $W$ bosons. 
The new physics scenario also includes the decay of a pair of electrically charged supersymmetric particles $\tilde{\chi}^\pm$ in two neutral supersymmetric particles $\tilde{\chi}^0 \tilde{\chi}^0$, undetectable and thus contributing to the missing transverse momentum, and two $W$ bosons. The dataset has 8 raw features and 10 high-level features. 

Unless specified differently,  we take $N(0)=10^5$ and $N_0=5\times 10^5$ and we vary the signal component in $N(S)\in[200,650]$.
We selected the following hyper-parameters:\\
$(M,\sigma,\lambda)=( 10^4,4.5,10^{-6})$ when using the raw features, increasing $\sigma$ to 5 when the high-level features are included.

The ideal significance is estimated by training a supervised classifier to discriminate between background and signal with a total of 2M examples,  following the approach in Ref.~\cite{Baldi:2014kfa}. The significance is then estimated by a cut-and-count analysis on the classifier output.

\paragraph{HIGGS} 
The HIGGS dataset \cite{Baldi:2014kfa} is made of simulated events in which the signal is given by the production of heavy Higgs bosons $H$. The final state is given by a pair of vector bosons $W^\pm W^\mp$ and two bottom quarks $b\bar{b}$ for both the reference and the signal components. The dataset has 21 raw feaures and 7 high-level feautures.

Unless specified differently, we choose $N(0)=10^5$, $N_0=5\times 10^5$ and we vary the signal component in $N(S)\in[1000,2500]$. 
We take the following hyper-parameters:\\
$(M,\sigma,\lambda)=( 10^4,7,10^{-6})$ when using the raw features and $\sigma=7.5$ when the high-level features are included.

The ideal significance is estimated as in the previous case by using the output of a supervised classifier trained to separate signal from background.

\subsection{Results}

\begin{figure}[H]
\centering
\begin{subfigure}{.5\textwidth}
  \centering
 \includegraphics[width=\textwidth]{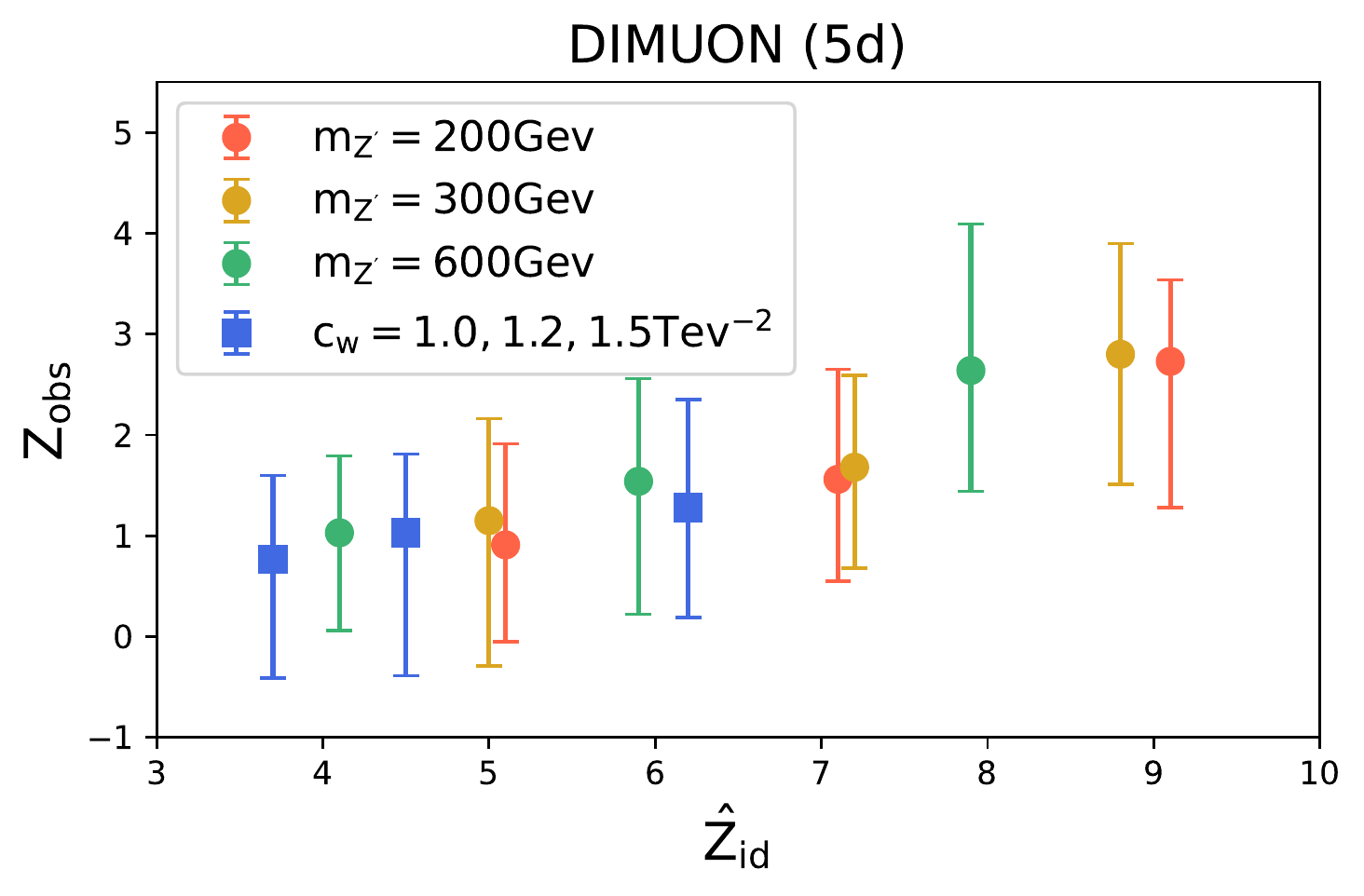}
 \caption{}\label{fig:Zobs_lowlevel_dimuon}
\end{subfigure}%
\begin{subfigure}{.5\textwidth}
  \centering
  \includegraphics[width=\textwidth]{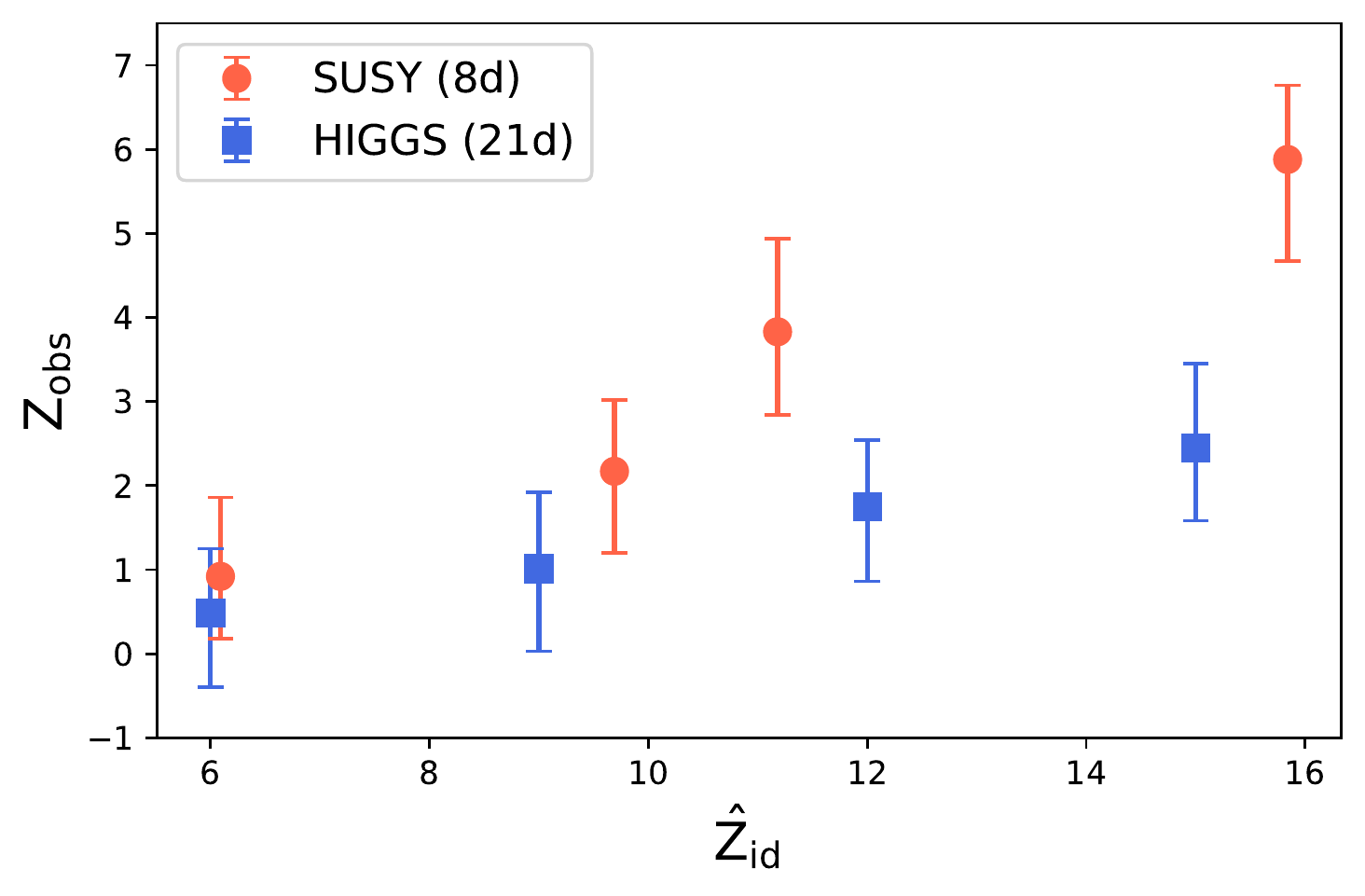}
  \caption{}\label{fig:Zobs_lowlevel_susyhiggs}
\end{subfigure}
\caption{Observed significance against estimated ideal significance with low-level input features.}
\label{fig:Zobs_lowlevel}
\end{figure}

\begin{figure}[H]
\centering
\includegraphics[scale=.45]{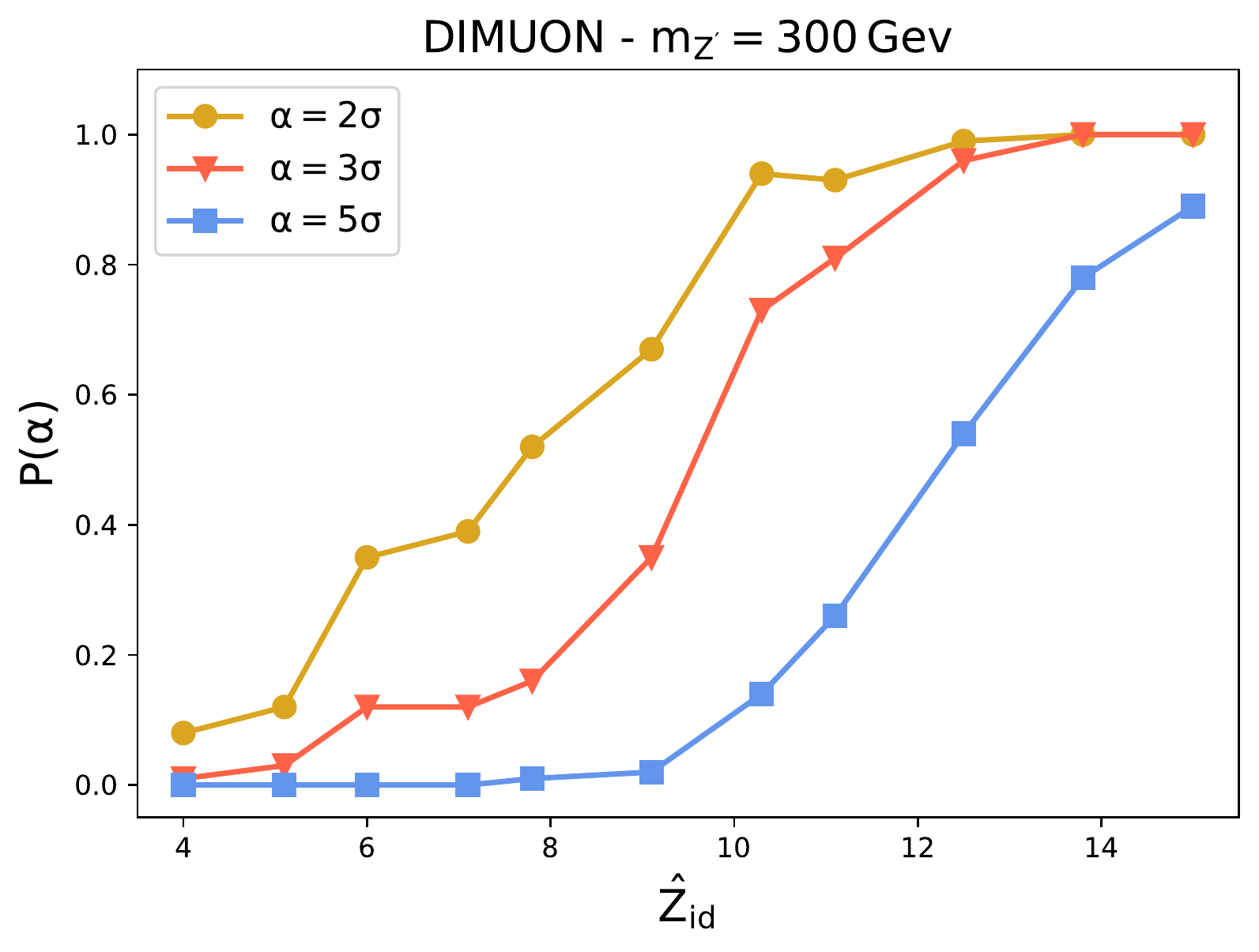}
\caption{Probability of finding a $\alpha = 2\sigma, 3\sigma, 5\sigma$ evidence for new physics as a function of the ideal significance.}
\label{fig:dilep_palpha}
\end{figure}

\begin{figure}[H]
\centering
\begin{subfigure}{.33\linewidth}
  \centering
 \includegraphics[width=\linewidth]{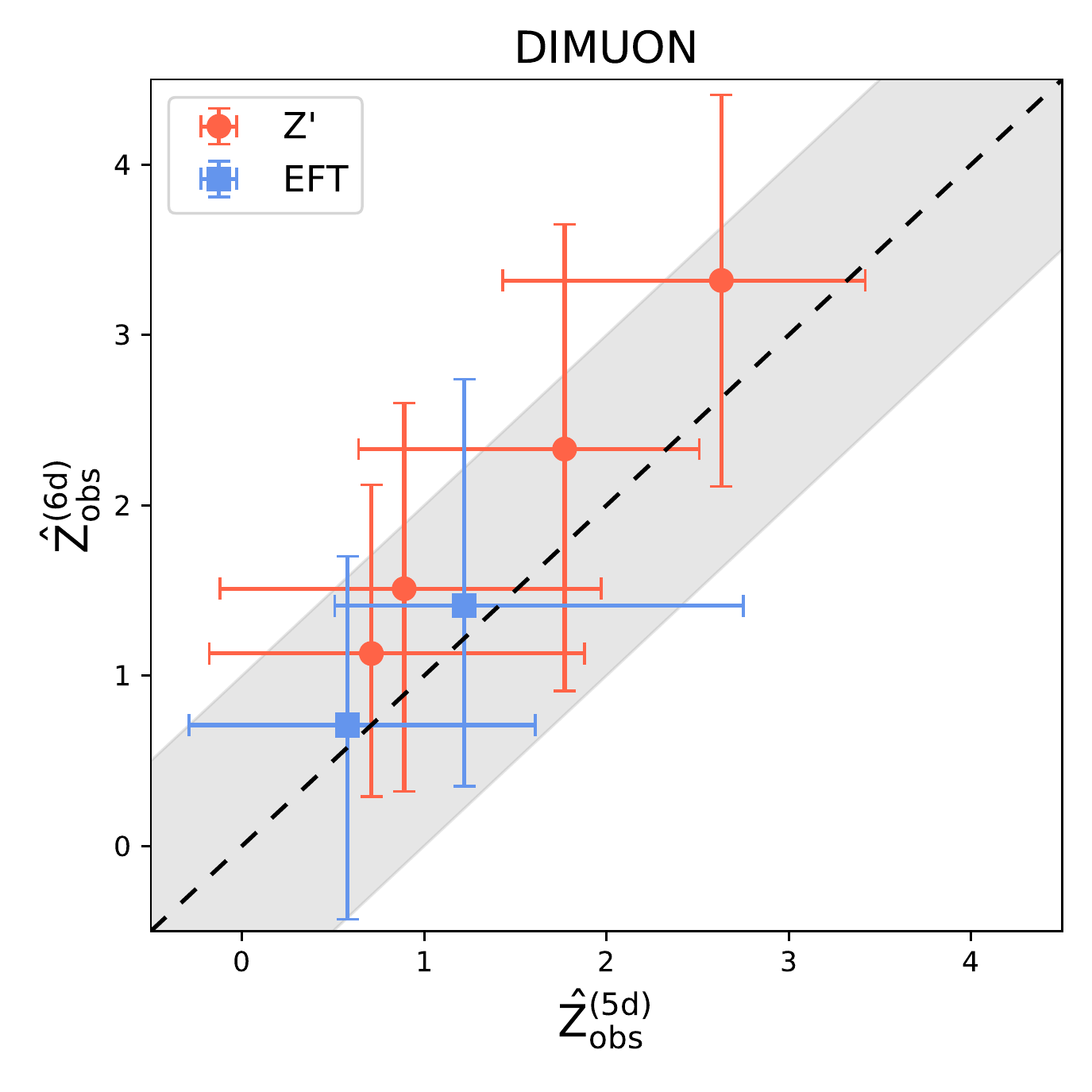}
\end{subfigure}%
\begin{subfigure}{.33\linewidth}
  \centering
 \includegraphics[width=\linewidth]{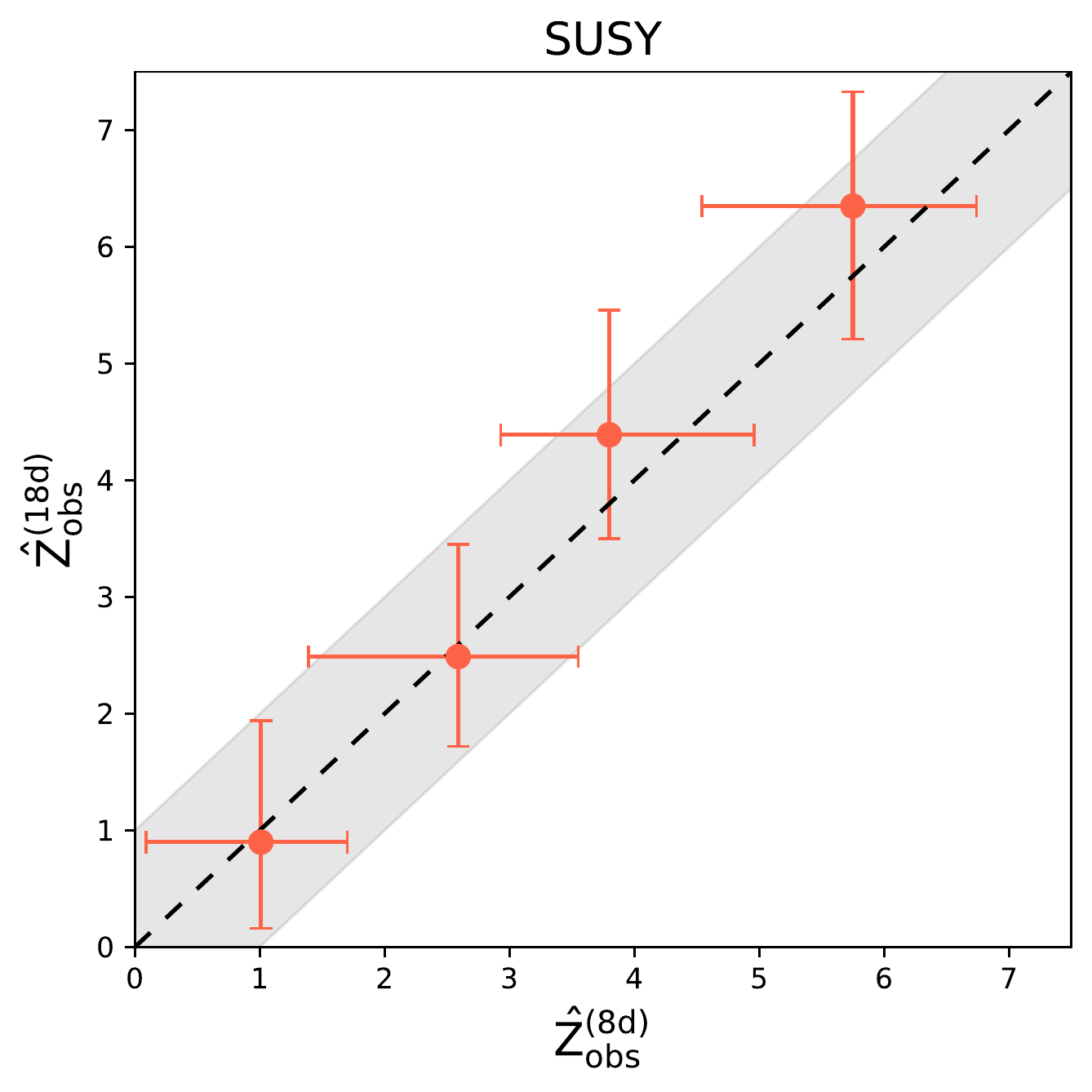}
\end{subfigure}%
\begin{subfigure}{.33\linewidth}
  \centering
  \includegraphics[width=\linewidth]{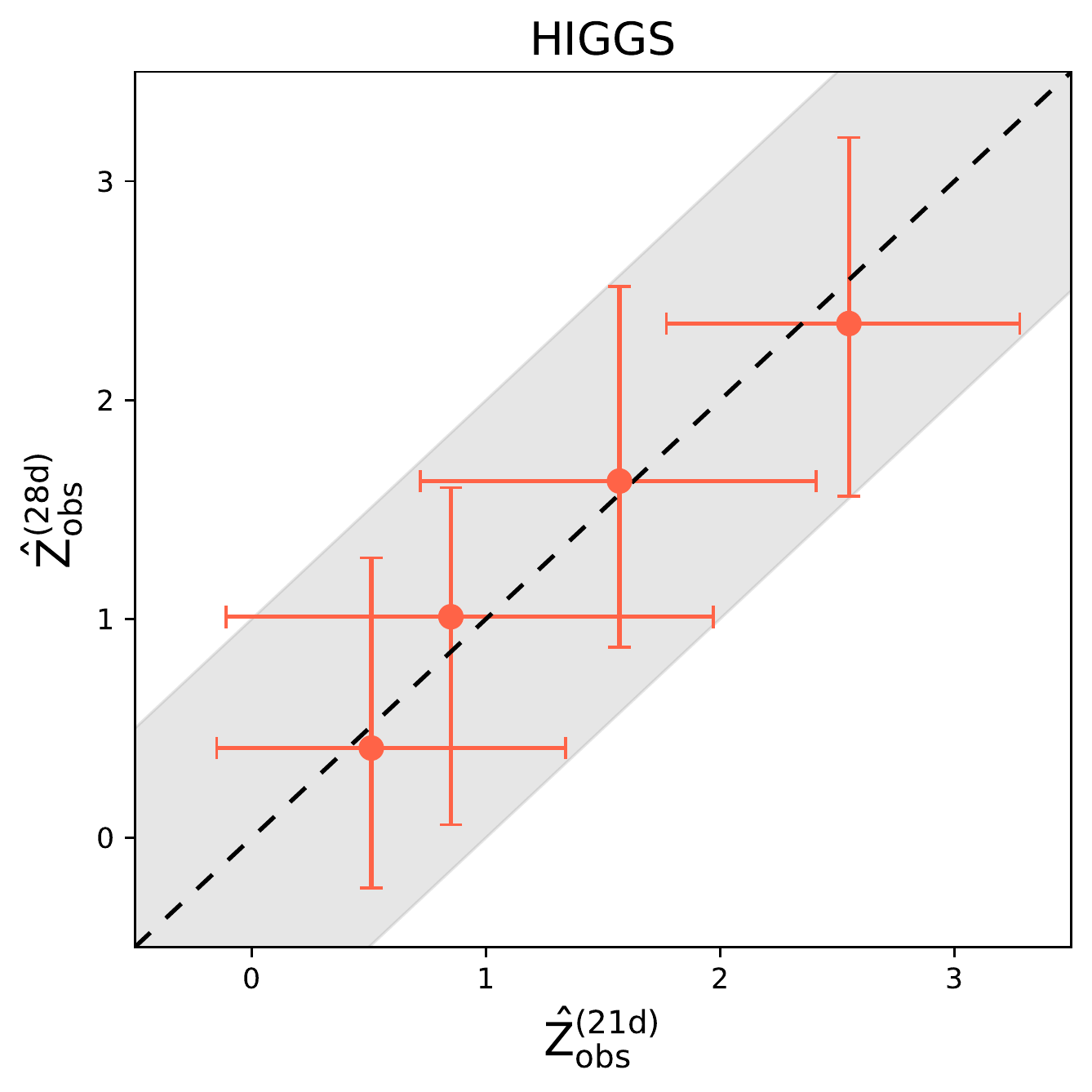}
\end{subfigure}
\caption{Comparison of the observed significance obtained with Falkon using low level features only and all the features. }
\label{fig:Zobs_all}
\end{figure}

\paragraph{Sensitivity to new physics} 
We discuss here the sensitivity of the model to the presence of new physics signals in the data.  
The test statistic distribution under the reference hypothesis is empirically reconstructed using 300 toy experiments while 100 toys are used to reconstruct the distribution of the test statistic under the alternative new physics scenarios. 
We show in Figures \ref{fig:Zobs_lowlevel} the median observed significance against the estimated ideal significance with Falkon trained on low-level features only.  These experiments were performed by varying the signal fraction $N(S)/N(0)$ (at fixed luminosity) and the type of signal (the latter in the DIMOUN case only). The error bars represent the  68\% confidence interval.  As expected for a model-independent strategy, the observed significance is always lower than what obtainable with a model-dependent approach.  The loss of sensitivity is more pronounced in higher dimensions.  Nevertheless,  we observe in all cases a correlation between the observed and the ideal significance.  In the DIMUON case,  the observe significance seems to depend weakly on the type of new physics signal.  
In Figure \ref{fig:dilep_palpha}, we show explicitly,  for the Z' new physics with $m_{Z'}=300$ Gev, the estimated probabilities to find a discrepancy of at least $\alpha$ for a given value of $\hat{Z}_{id}$.  Similar results are obtained with the other types of signal.  To test the ability of the kernel-based approach to extract useful information from data,  we show in Figure \ref{fig:Zobs_all} that adding the high-level features does not significantly improve the results,  especially in higher dimensions.  The plot includes the observed significance,  with the bar showing the 68\% confidence interval and the grey area representing the region $Z^{(\textrm{all})}_{obs}=Z^{(\textrm{low-level})}_{obs} \pm \sigma$.

\begin{figure}[H]
\centering
\begin{subfigure}{.49\linewidth}
  \includegraphics[width=\linewidth]{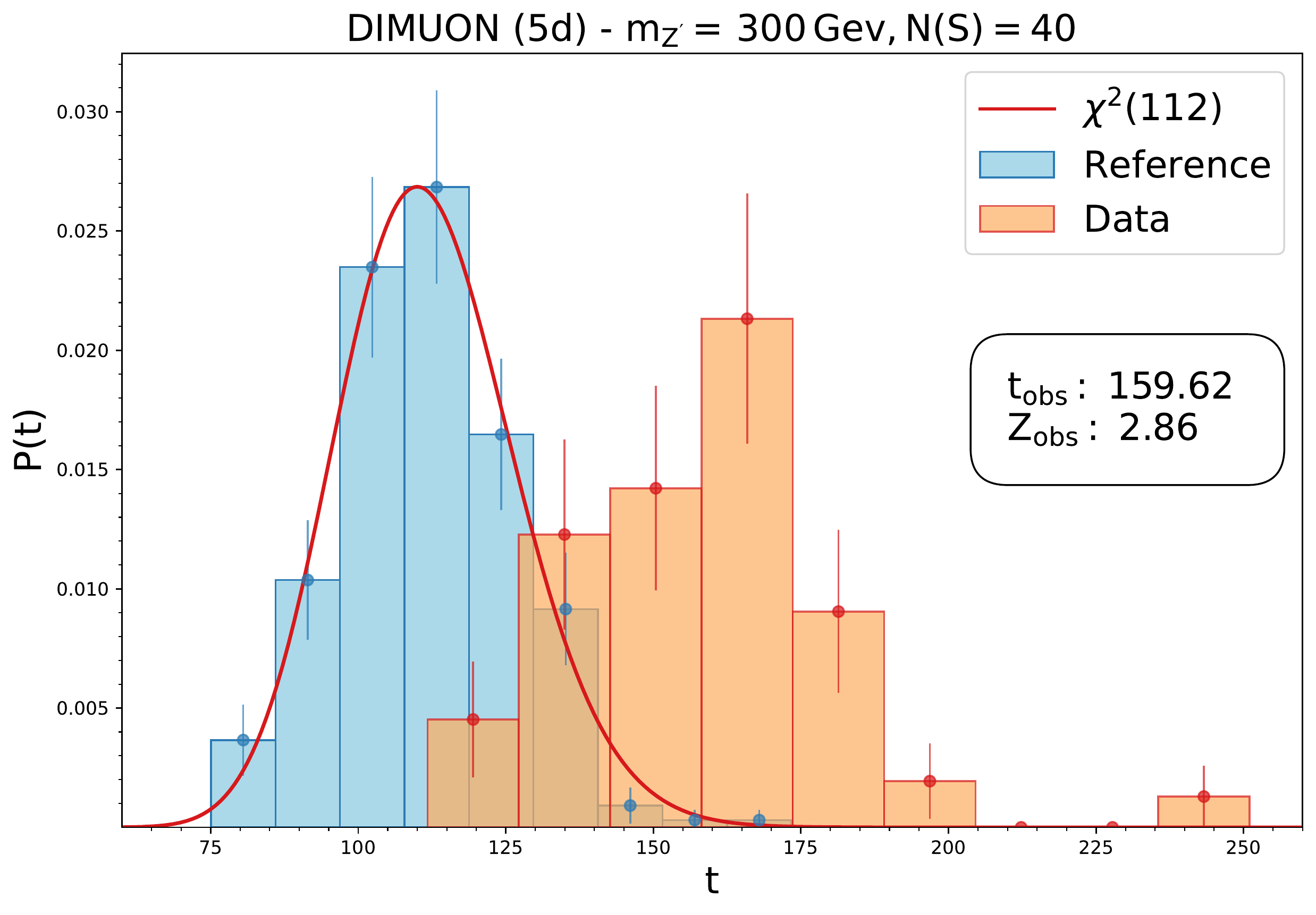}
\end{subfigure}
\centering
\begin{subfigure}{.49\textwidth}
  \includegraphics[width=\textwidth]{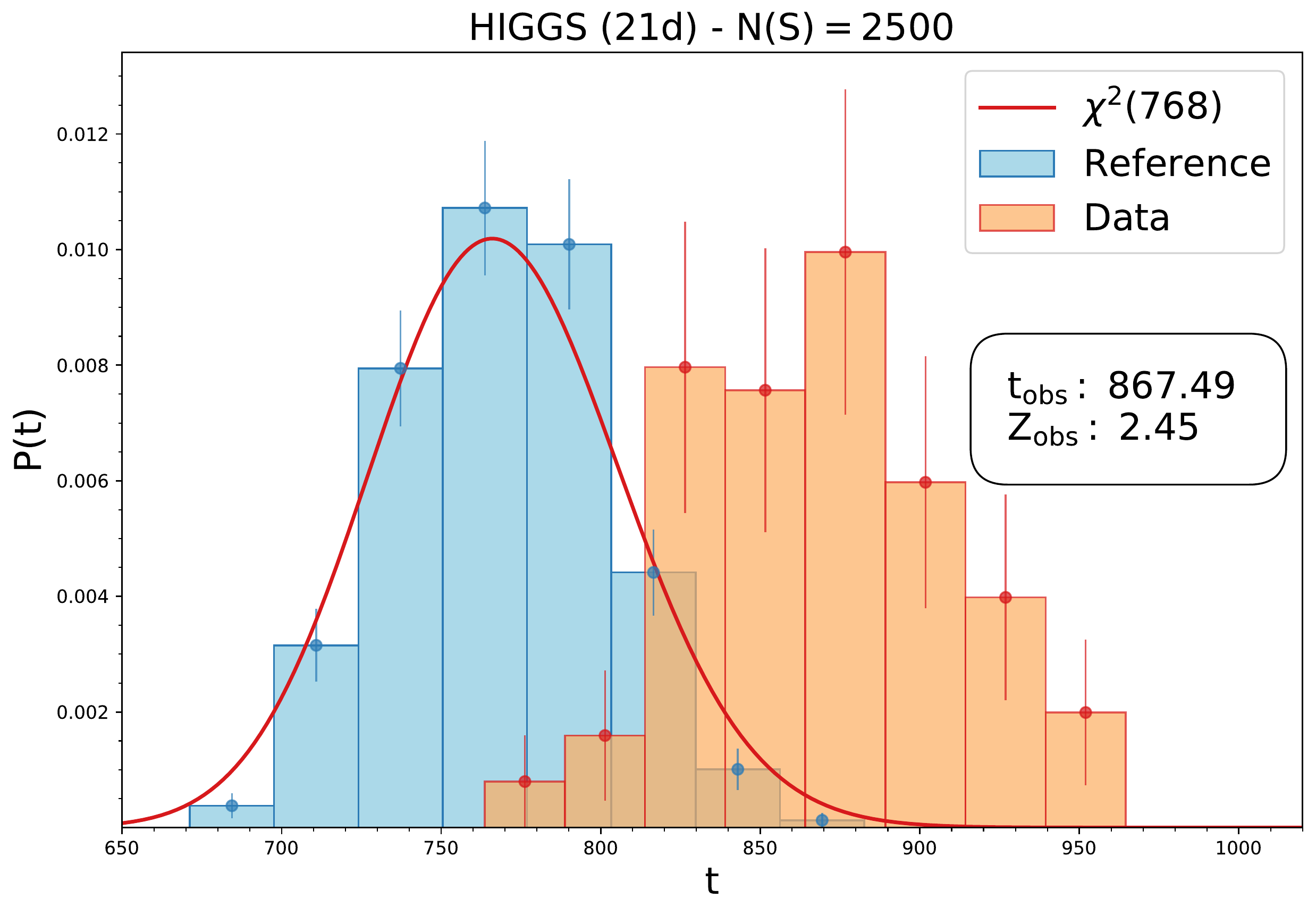}
\end{subfigure}
\caption{Distribution of the test statistics under the null and alternative hypotheses for the DIMUON (left) and HIGGS (right) datasets.}
\label{fig:exp_refsig}
\end{figure}

\begin{figure}[H]
\centering
\begin{subfigure}{.33\linewidth}
  \centering
 \includegraphics[width=\linewidth]{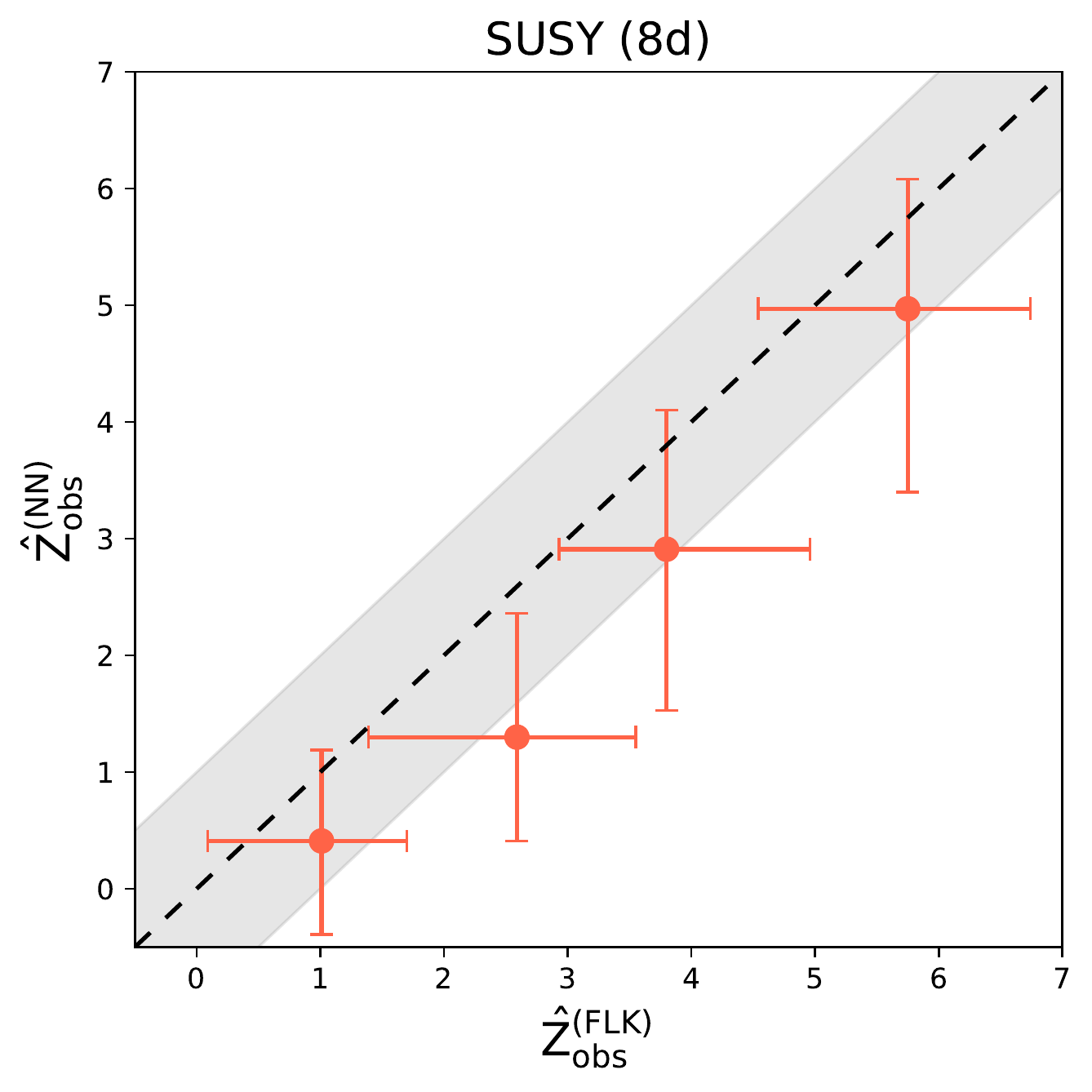}
\end{subfigure}%
\begin{subfigure}{.33\linewidth}
  \centering
  \includegraphics[width=\linewidth]{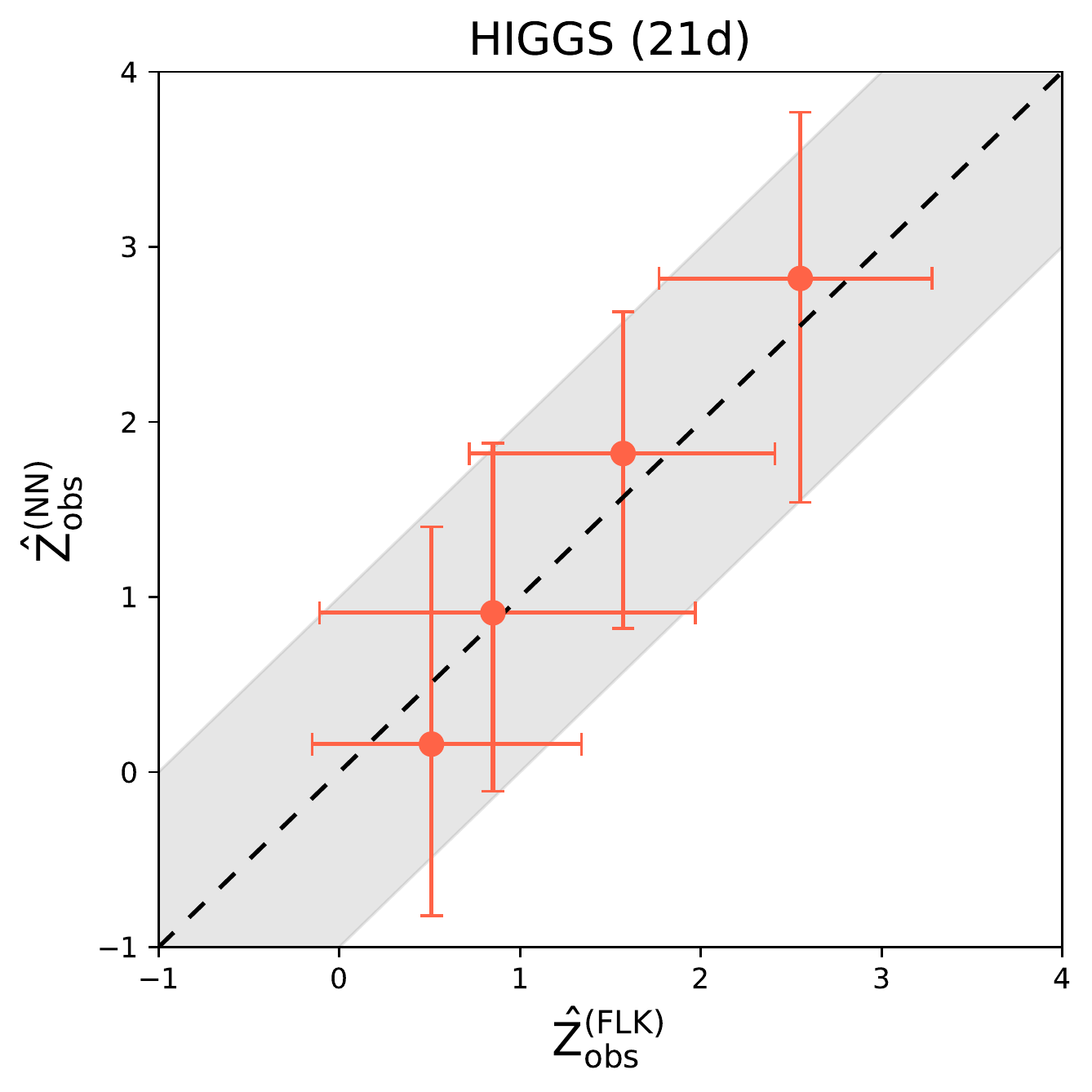}
\end{subfigure}
\caption{Observed significance with the Falkon implementation against neural networks.}
\label{fig:Zobs_higgs_susy_NN}
\end{figure}

\paragraph{Comparison with neural networks} 
To compare the kernel-based approach with the neural network implementation,  we considered the results from Ref.~\cite{DAgnolo:2019vbw} for the DIMUON dataset,  while we trained the latter on the SUSY and the HIGGS datasets.  The considered neural network has 2 hidden layer with 10 neurons each and a weight clipping of $w_{\text{clip}}=0.87$ for SUSY,  while it has 5 layers with 6 neurons each layer and $w_{\text{clip}}=0.65$ for HIGGS.  Training is stopped after $3\times 10^5$ epochs.  The results are summarized in Figure \ref{fig:Zobs_higgs_susy_NN}.
We see that,  overall,  the two approaches give similar results and the degradation of the sensitivity in high dimensions affects both.  
We notice that in the DIMUON case,  the kernel approach is slightly less sensitive,  as it can be seen from the results presented in Section 5 of Ref.~\cite{DAgnolo:2019vbw} against Figures \ref{fig:Zobs_lowlevel_dimuon} and \ref{fig:dilep_palpha}.  On the other hand,  by looking at Figure \ref{fig:Zobs_higgs_susy_NN} we see that,  for the HIGGS dataset,  the kernel approach gives a higher observed significance while,  for the SUSY dataset,  the two methods give almost identical results.
On the other hand,  the average training times,  summarized in Table \ref{table:tr_times},  demonstrate an advantage in favor of the kernel approach of orders of magnitude.  This also allows efficient training on single GPU machines and ensures high scalability for multi-GPU systems,  as shown in Ref.~\cite{falkonlibrary2020}.  

\begin{table}[H]
\centering
\begin{tabular}{l | lll}
 \multicolumn{4}{c}{} \\
 \toprule
 Model & DIMUON & SUSY & HIGGS \\
 \midrule
 \bf{FLK} & \bf{(44.9 $\pm$ 3.4) s} & \bf{(18.2 $\pm$ 1.2) s} & \bf{(22.7 $\pm$ 0.4) s} \\
 NN & (4.23 $\pm$ 0.73) h & (73.1 $\pm$ 10) h & (112 $\pm$ 9) h \\
 \bottomrule
 \end{tabular}
  \caption{Average training times per single run with standard deviations (low level features and reference toys).  Nota that time measured in hours (for NN) and seconds (for Falkon).}
 \label{table:tr_times}
\end{table}

\begin{figure}[H]
\centering
\begin{subfigure}{.45\linewidth}
\centering
\includegraphics[width=\linewidth]{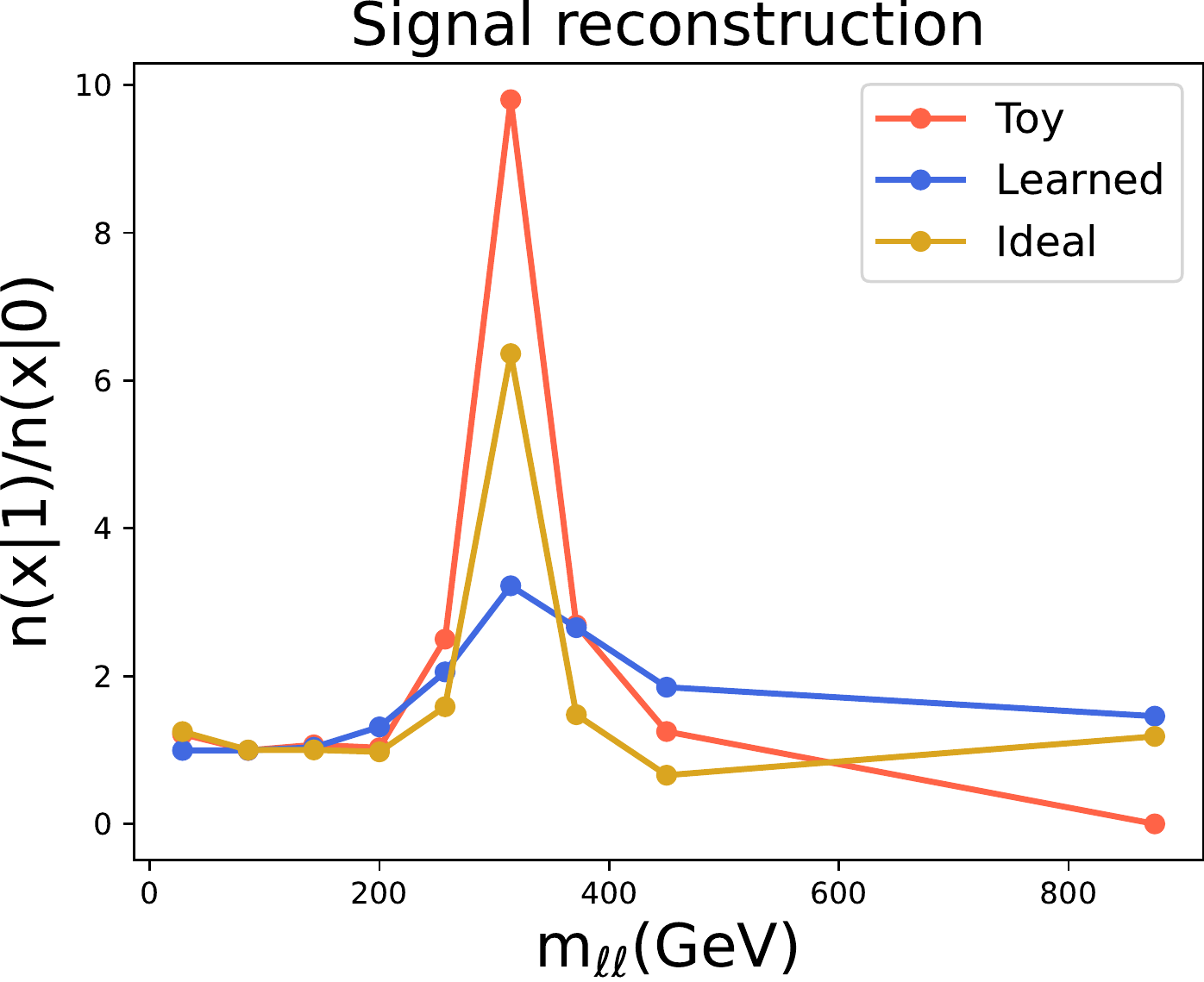}
\end{subfigure}
\hfill
\begin{subfigure}{.45\linewidth}
\centering
\includegraphics[width=\linewidth]{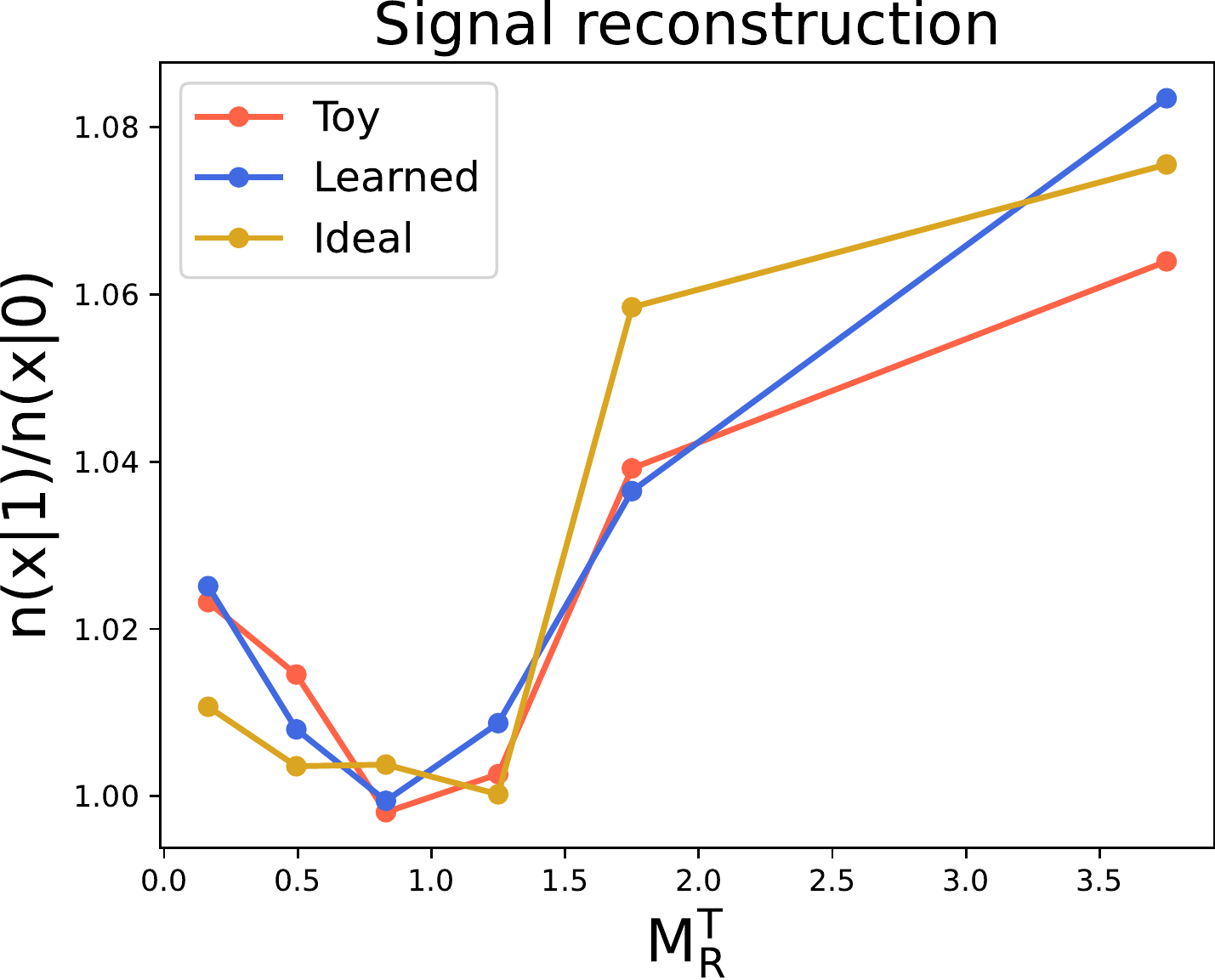}
\end{subfigure}
\caption{Examples of reconstructed density ratios as a functions of high-level features (not given as inputs) for the DIMUON (left) and SUSY (right) datasets with new physics components in the data.  Note that the SUSY dataset is normalized.}
\label{fig:lrplot}
\end{figure}

\paragraph{Learned density ratio} 
As discussed in Section \ref{sec:framework}, the function approximated by using the weighted cross-entropy loss is the density ratio given in Eq.\eqref{nratio}.  The latter can be directly inspected to characterize the nature of the ``anomalies" in the experimental data,  if found significant.  We report in Figure \ref{fig:lrplot} examples of the reconstructed density ratios as functions of certain high-level features (not given as inputs) together with estimates of the true ratios and extrapolations from the data used for training. 
The learned density ratio is constructed by re-weighting the relevant high-level feature of the reference sample by $e^{f_{\hat{w}}(x)}$ (evaluated on the reference training data),  binning it and taking the ratio with the same binned reference sample (unweighted).  The toy density ratio is computed by replacing the numerator with the binned distribution of the high-level feature of the toy data sample.  The ideal case is obtained in the same way but using a large ($\geq$1M) data sample instead. 

\begin{figure}[H]
\begin{subfigure}{.49\linewidth}
\centering
 \includegraphics[width=\linewidth]{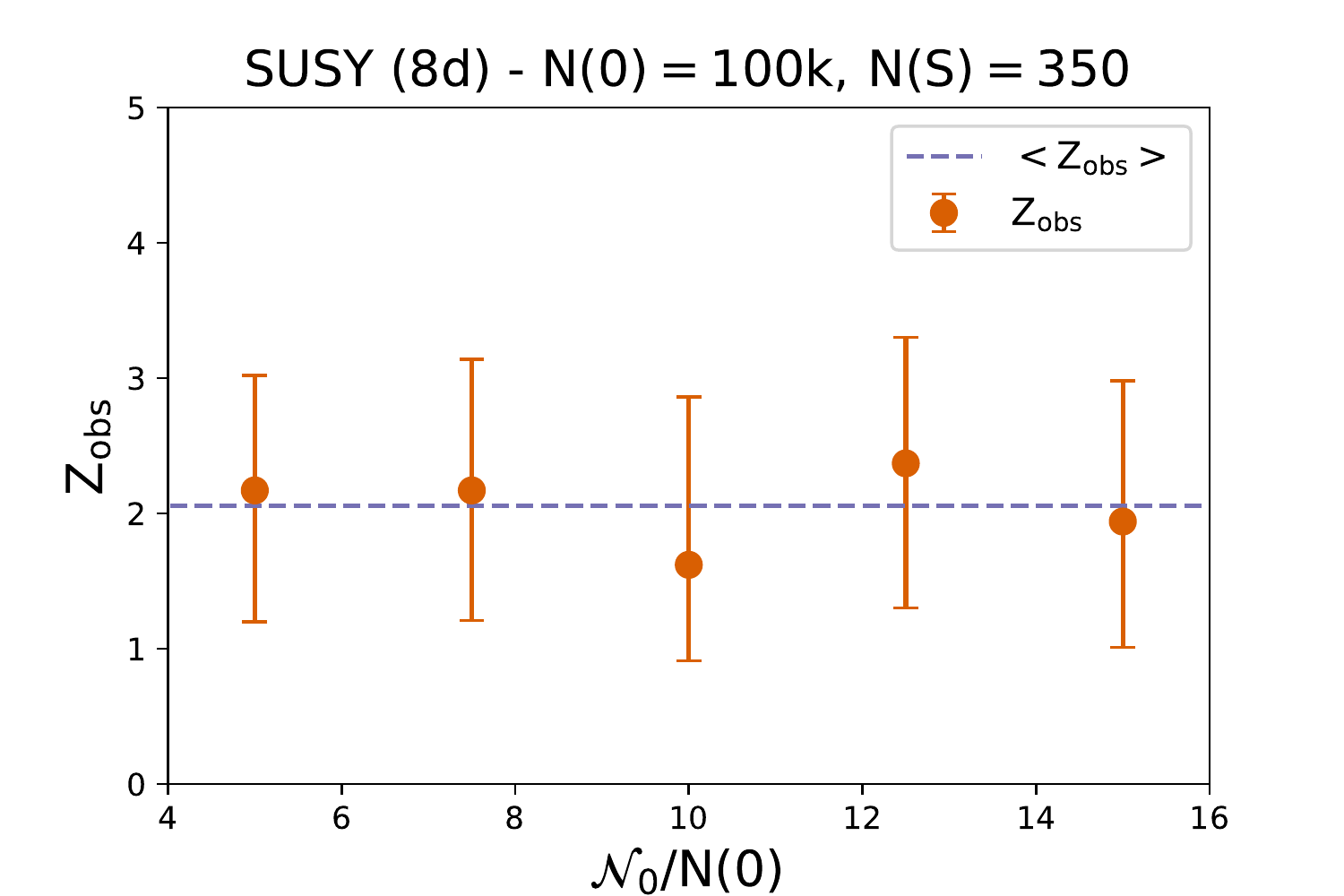}
 \caption{}\label{fig:ZobsvsN0}
 \end{subfigure}
 \begin{subfigure}{.49\linewidth}
\centering
\includegraphics[width=\linewidth]{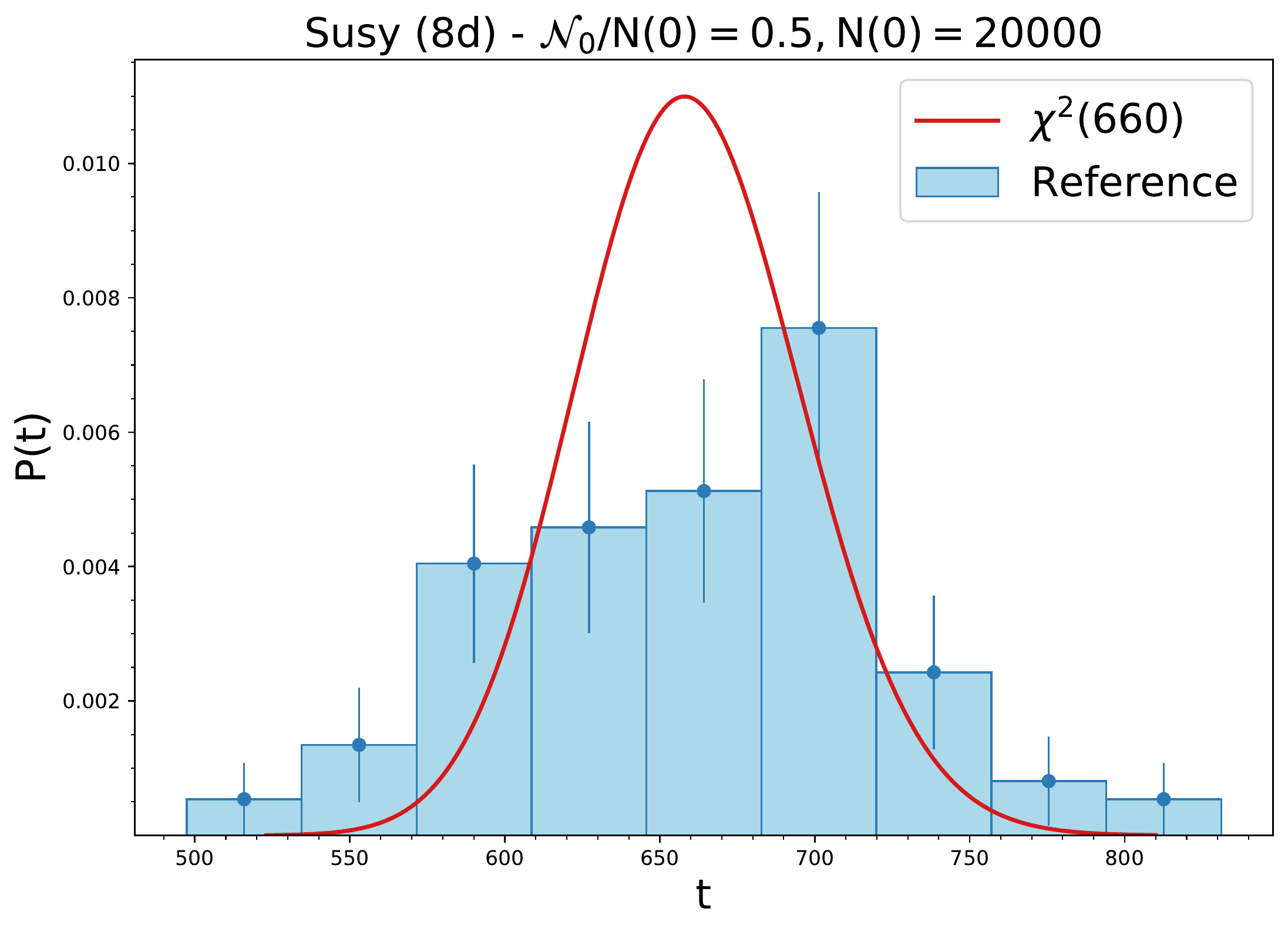}
\caption{}\label{fig:ref50k}
 \end{subfigure}
\caption{Observed significance as a function of the size of the reference sample(left). Example of distribution of the test statistics given a small reference sample (right).}
\end{figure}

\paragraph{Size of the reference sample} 
A larger reference sample yields a better representation of the reference model,  which is crucial for a model-independent search.  In Figure \ref{fig:ZobsvsN0}, we see that as long as $\mathN_0/N(0)\gtrsim 1$, the median observed significance is indeed stable.  On the other hand, when the reference sample is too small ($\mathN_0/N(0)< 1$), we observe that the correspondence between the distribution of the test statistics and the $\chi^2$ distribution breaks down, see Figure \ref{fig:ref50k}.  We observe this behavior for all the datasets.  Then,  it is in general a good approach to take a reference sample as large as possible keeping in consideration the computational cost of training on a possibly very large dataset. 

\paragraph{Resources}
The models based on Falkon have been trained on a server with the specifications reported in Table \ref{table:specs}. The NN experiments have been performed on a CPU farm with 32 computing nodes of Intel 64 bit dual processors,  for a total amount of 712 core. 

 \begin{table}[H]
  \centering
  \begin{tabular}{l | l}
  \multicolumn{2}{c}{} \\
  \toprule
    OS & Ubuntu 18.04.1 \\
    CPU(s) & $2 \times$ Intel(R) Xeon(R) Silver 4116 CPU\\
    RAM & $256$GB\\
    GPU(s) & $2 \times$ NVIDIA Titan Xp (12 GB RAM)\\
    CUDA version & $10.2$\\
    \bottomrule
  \end{tabular}
   \caption{Specifications of the machine used to perform the experiments with Falkon.}
   \label{table:specs}
\end{table}

%%%%%%%%%

\section{Conclusions}\label{sec:conclusions}
In this work we have presented a machine learning approach for model-independent searches applying kernel-based machine learning models to the ideas introduced in Ref.~\cite{DAgnolo:2018cun,DAgnolo:2019vbw}.  Our approach is powered at its core by Falkon,  a recent library developed for large scale applications of kernel methods. The focus of our work is on computational efficiency.  Indeed,  the original neural network proposal suffers from long training times which,  combined with a toy-based hypothesis testing framework, makes the use of the algorithm challenging in high dimensional cases.  Our model delivers comparable performances with a dramatic reduction in training times, as shown in Table \ref{table:tr_times}.  As a consequence,  the model can be efficiently trained on single GPU machines while possessing high scalability for multi-GPU systems \cite{falkonlibrary2020}. In contrast,  the neural network implementation crucially relies on per toy parallelization,  hence the need for large scale resources such as CPU/GPU clusters.

Similarly to Ref.~\cite{DAgnolo:2019vbw}, the applicability of the proposed method relies on a heuristic procedure to tune the algorithm hyper-parameters. A more in-depth understanding of the interplay between the expressibility of the model, its complexity and the topology of the input dataset could lead to more performant and better motivated alternatives to the current hyper-parameter selection. Further investigations are left for future work. 
One possibility would be to find a more principled way to relate Falkon hyper-parameters to physical quantities.  This could also allow the introduction of explicit quantities to be optimized,  opening to the possibility of  applying modern optimization techniques for the selection of the hyper-parameters. 

Besides the challenges related to the algorithm optimization and regularization, an essential development for the application to realistic data analysis concerns the treatment of systematic uncertainties which has not been considered in the present work.
 This aspect was successfully addressed on a recent work \cite{dAgnolo:2021aun} in the context of the neural network implementation. 

Finally,  the boost in efficiency provided by the model developed in this work could extend the landscape of applicability of this analysis strategy to other use cases, beyond the search for new physics, and to other domains.  In particular, the application to multivariate data quality monitoring in real time  is currently under study.

%%%%%%%

$$$$
\bf Acknowledgements: \rm Lorenzo Rosasco acknowledges the financial support of the European Research Council (grant SLING 819789),  the AFOSR projects FA9550-18-1-7009, FA9550-17-1-0390 and BAA-AFRL-AFOSR-2016-0007 (European Office of Aerospace Research and Development), the EU H2020-MSCA-RISE project NoMADS - DLV-777826, and the Center for Brains, Minds and Machines (CBMM), funded by NSF STC award CCF-1231216. We gratefully acknowledge the support of NVIDIA Corporation for the donation of the Titan Xp GPUs and the Tesla k40 GPU used for this research. M.P.  and G.G.  are supported by the European Research Council (ERC) under the European
Union’s Horizon 2020 research and innovation program (grant agreement no 772369).  A.W. acknowledges support from the Swiss National Science Foundation under contract 200021-178999 and PRIN grant 2017FMJFMW.

\clearpage

\bibliographystyle{unsrt}

\bibliography{references}

\clearpage

\appendix
\appendixpage
\addappheadtotoc

\section{Loss functions and target functions}\label{app:SLT}
Different loss functions determine different goals via an associated target function $f^*$.  This is the function learned by the model in the large-sample limit and it can be computed, given a loss function $\ell(y,f(x))$,  by considering the expected risk
\be
L(f)=\int \ell(y,f(x)) \,dp(x,y),
\ee
where $p(x,y)$ is the true joint distribution.  It can be further rewritten as
\be
\bal
L(f)&=\int \ell(y,f(x)) p(x,y) \,dx dy \\
&=\int p(x)\, dx \int \ell(y,f(x)) \,p(y|x) \,dy.
\eal
\ee
One can then find the minimizer simply as
\be 
f^*=\argmin_{f\in\mathbb{R}}\int \ell(y,f(x)) \,p(y|x) \,dy,\quad \forall x,
\ee
with $p(x) p(y|x)=p(x,y)$. In the case of the weighted cross-entropy loss,  one has
\be
\ell(y,f(x))=a_0\,(1-y)\log\left(1+e^{f(x)}\right)+a_1\,y\log \left(1+e^{-f(x)}\right),
\ee
with $y=\{0,1\}$.  One then simply takes the derivative and sets it equal to zero,  obtaining the following minimizer
\be
f^*=\log\frac{p(1|x)}{p(0|x)}\frac{a_1}{a_0}.
\ee

%%%%%%%%%%%

\section{Falkon}\label{app:falkon}
In this appendix,  we provide more details on Falkon~\cite{falkonlibrary2020},  the algorithm powering our model.  The original library includes an implementation based on the square loss, which we do not discuss here.
The core ideas from a theoretical and algorithmic viewpoint are developed in Ref.~\cite{rudi2017falkon,marteauferey2019globally,marteauferey2019leastsquares}. 

The problem of minimizing the regularized empirical risk in Eq.\eqref{reg_ERM} is formulated in terms of an approximate Newton method (see Algorithm 2 of Ref.~\cite{falkonlibrary2020})
The model is based on the Nystr\"{o}m approximation,  which is used twice.  First to reduce the size of the problem,  by considering solutions of the form shown in Eq.\eqref{kernel_sol}.  Then,  it is again used to derive an approximate Newton step.  
At every step,  preconditioned conjugate gradient descent is run for a limited number of iterations with a decreasing sequence of regularization parameters $\lambda_k$, down to the desired regularization level.  We choose $k=1$ in our experiments,  as we did not observe any benefit in selecting more values. 
The preconditioner plays here the role of approximate Hessian. 
Given $(x_i,y_i)_{i=1}^m$ selected uniformly at random from the dataset and let $T$ be the Cholesky decomposition of $K_{mm}$ then
the approximate Hessian $\tilde{H}$ has the form
\begin{equation}\label{apprH}
  \tilde{H} = \frac{1}{m}T\tilde{D}_kT^\intercal + \lambda_k I,
\end{equation}
where $\tilde{D}_k \in \mathbb{R}^{m \times m}$ is a diagonal matrix s.t. the $i$-th element is the second derivative of the loss $\ell^{\prime \prime}(y_i,f(x_i),)$ with respect to the first variable. To preserve efficiency, this matrix is never built explicitly but we build it in terms of Cholesky decomposition: let $A$ be the Cholesky decomposition of Eq.\eqref{apprH}, we compute
\begin{equation*}
  P = T^{-1}A^{-1} \qquad \tilde{H}^{-1} = PP^\intercal.
\end{equation*}  
Then conjugate gradient is applied to solve the preconditioned problem at time $k$
\begin{equation*}
  P^\intercal (K_{nm}^\intercal D_k K_{nm} + \lambda_k I) P \beta = P^\intercal K_{nm}^\intercal g_k,
\end{equation*}
where $g_k \in \mathbb{R}^n$ such that $(g_k)_i = l^{\prime}(f(x_i), y_i)$.  With this strategy,  the overall computational cost to achieve optimal statistical bounds is $\mathcal{O}(n\sqrt{n} \log n)$ in time,  and in $\mathcal{O}(n)$ in memory,  making it suitable for large scale problems.  

%%%%%%%%%%

\section{1D example}\label{app:univ}
We consider here a simple univariate toy scenario taken from Ref.~\cite{DAgnolo:2018cun}.  We use this example to present explicitly all the steps discussed in Section \ref{sec:framework} and Section \ref{sec:model}.

\begin{figure}[H]
\centering
\begin{subfigure}{0.32\textwidth}
\centering
\includegraphics[width=1.1\textwidth]{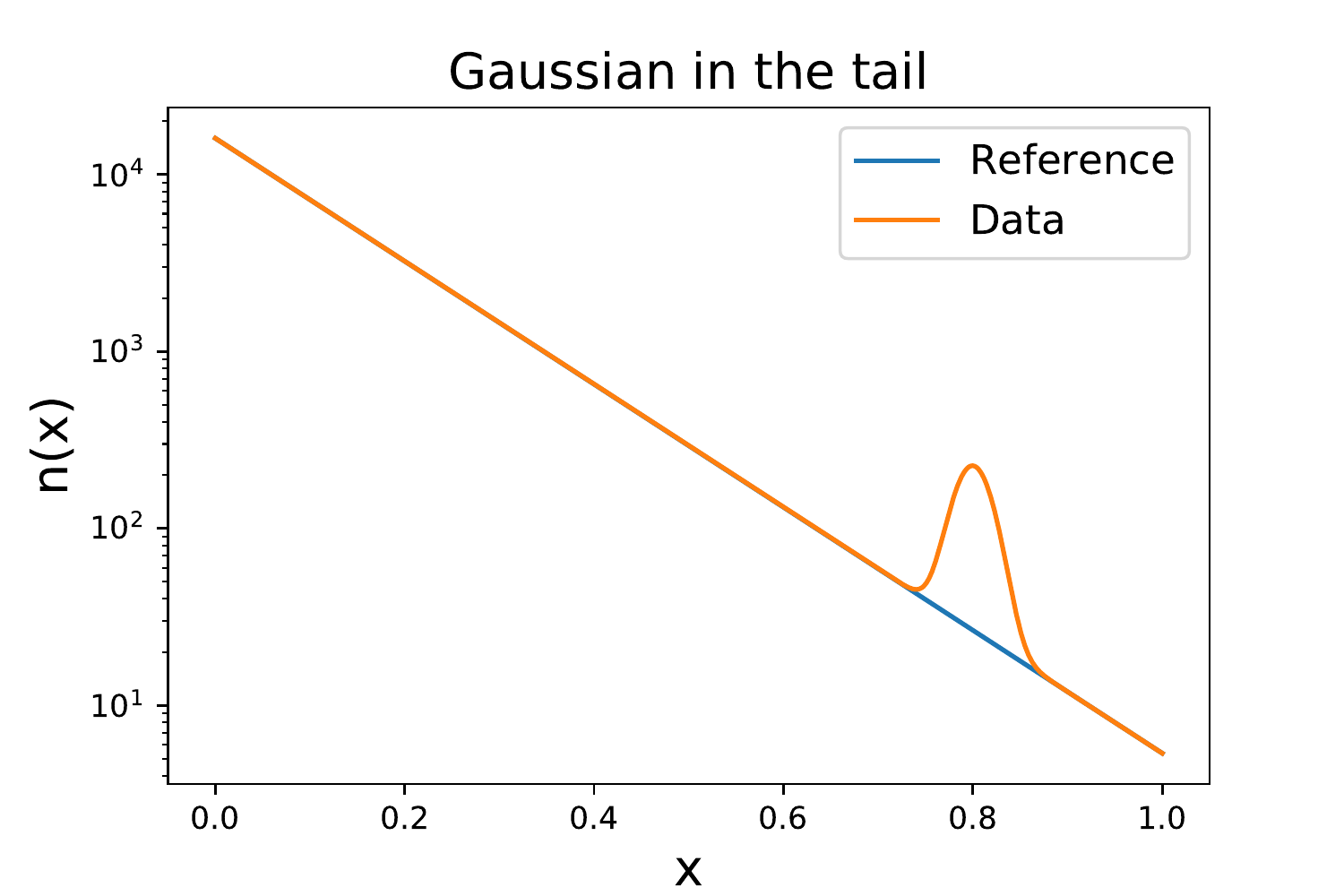}
\end{subfigure}
\begin{subfigure}{0.32\textwidth}
\centering
\includegraphics[width=1.1\textwidth]{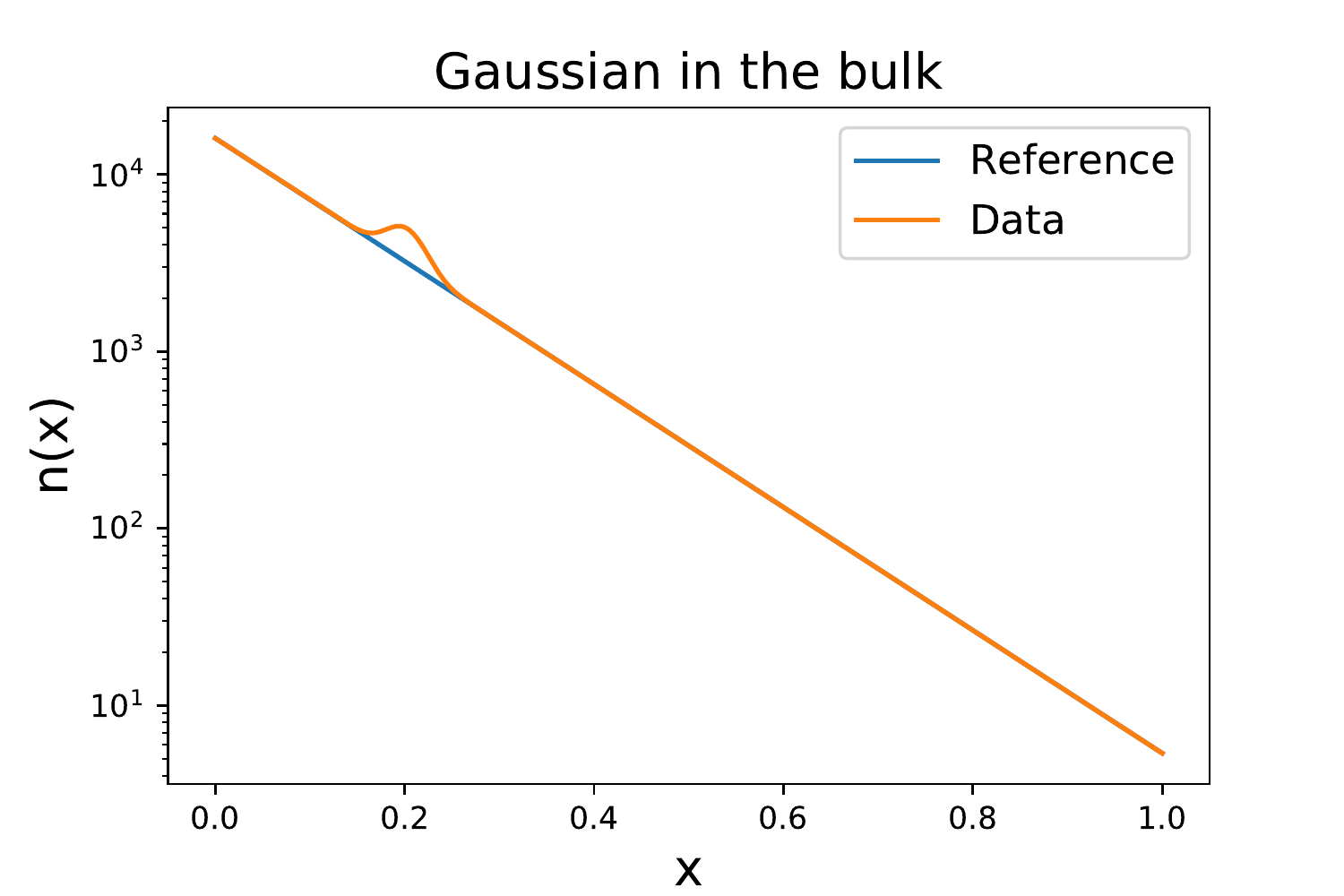}
\end{subfigure}
\begin{subfigure}{0.32\textwidth}
\centering
\includegraphics[width=1.1\textwidth]{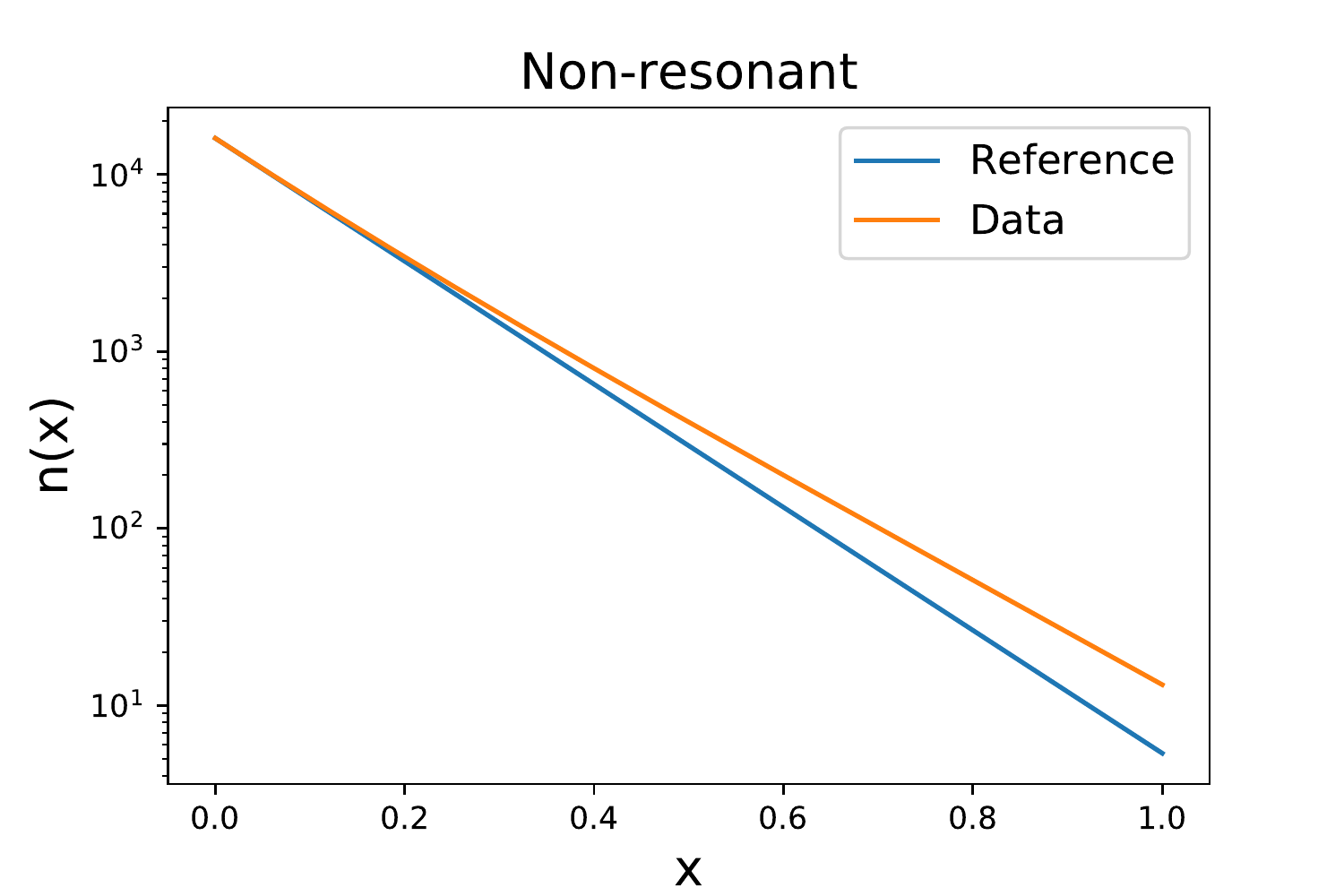}
\end{subfigure}
\caption{True univariate densities.}
\label{fig:univ_distr}
\end{figure}

\paragraph{Data.} We know here explicitly both the reference and the true data generating distributions. 
The former is an exponential 
\be\label{univ_ref}
n(x|0)=N(0)\,8\,e^{-8x}.
\ee
The latter is given by the reference distribution combined with a signal component and reads as
\be\label{univ_n}
\bal
n(x|1) &= n(x|0)+n(x|S) \\
&= N(0) p(x|0) + N(S) p(x|S),
\eal
\ee
where $p(x|S)$ is the distribution of the signal alone and $N(S)$ is the expected number of signal events. 
We consider three types of signals, two that are localized in a region of the input feature (resonant) and one that is not (non-resonant). They are given by the following expressions (see also Figure \ref{fig:univ_distr}): 
\begin{itemize}
\item A Gaussian distribution centered in the tail of the exponential background
\be
\bal
&n(x|S_1)=N(S_1)\frac{1}{\sqrt{2\pi}\sigma} e^{(x-\mu_1)^2/2\sigma^2},\\
&\mu_1=0.8,\quad \sigma=0.02,\quad N(S_1)=10,\eal
\ee
\item A Gaussian distribution centered in the bulk of the exponential background
\be
\bal
&n(x|S_2)=N(S_2)\frac{1}{\sqrt{2\pi}\sigma} e^{(x-\mu_2)^2/2\sigma^2},\\
& \mu_2=0.2,\quad \sigma=0.02,\quad N(S_2)=90,
\eal
\ee
\item A non-resonant signal given by
\be
n(x|S_3)=N(S_3)\,256\,x^2 e^{-8 x},\quad N(S_3)=90.
\ee
\end{itemize}

We select a large reference sample of size $\mathN_0=2\times 10^5$  and an expected number of background events $N(0)=2\times 10^3$.  The size of the data sample is then $\mathN_1\sim \textrm{Pois}(N(y))$, with $N(1)=N(0)+N(S)$ when the data sample is generated according to the true data distribution.

\begin{figure}[H]
\centering
    \includegraphics[scale=.5]{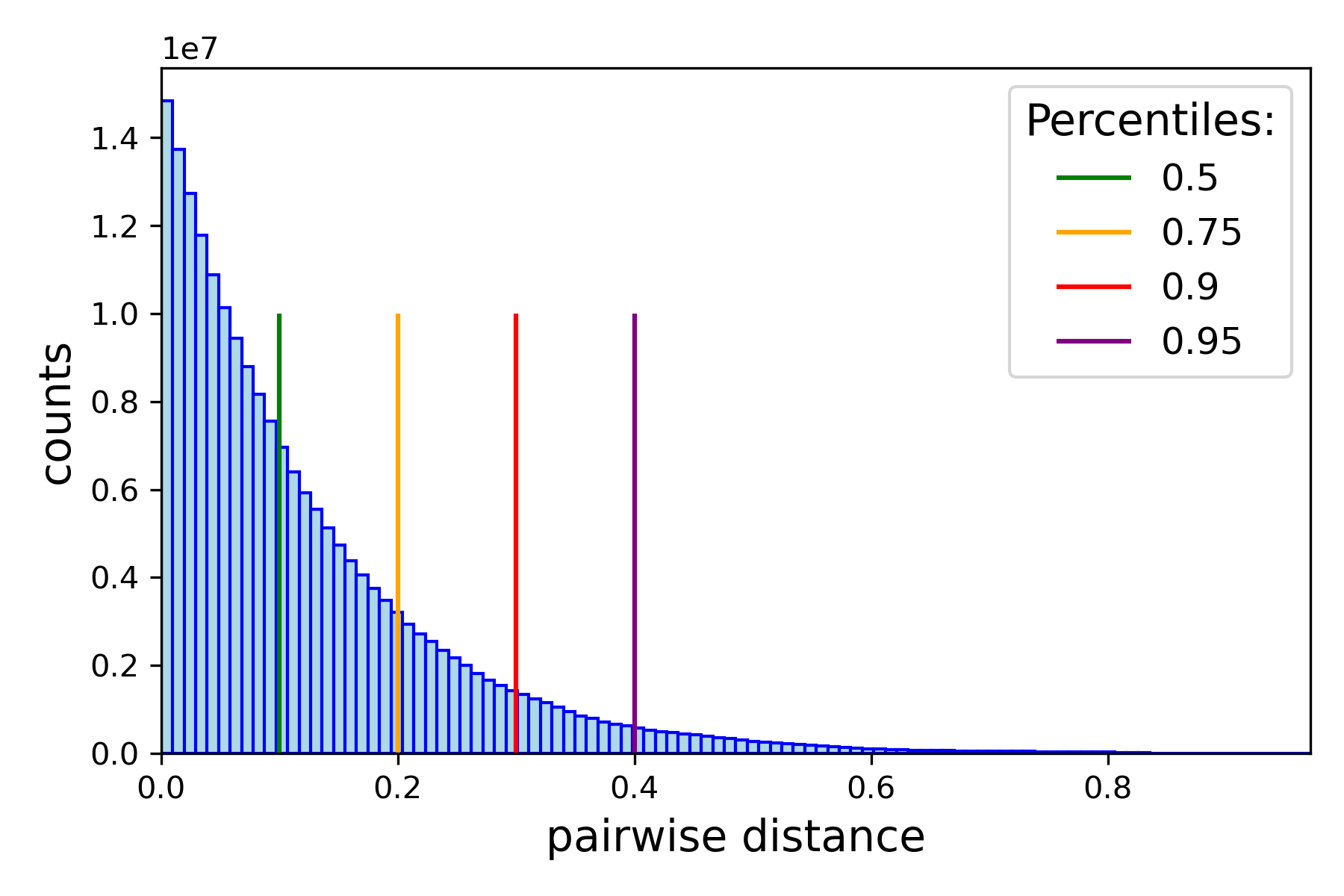}
    \caption{Euclidean pairwise distance. }
    \label{fig:pariw_univ}
\end{figure}

\paragraph{Model tuning}  In Figure \ref{fig:pariw_univ} we show the distribution of the pairwise distances from which we select the bandwidth as approximately the 90th percentile.  In this case,  it corresponds to $\sigma\approx 0.3$.

\begin{figure}[H]
\centering
\begin{subfigure}{0.49\linewidth}
\centering
    \includegraphics[width=\linewidth]{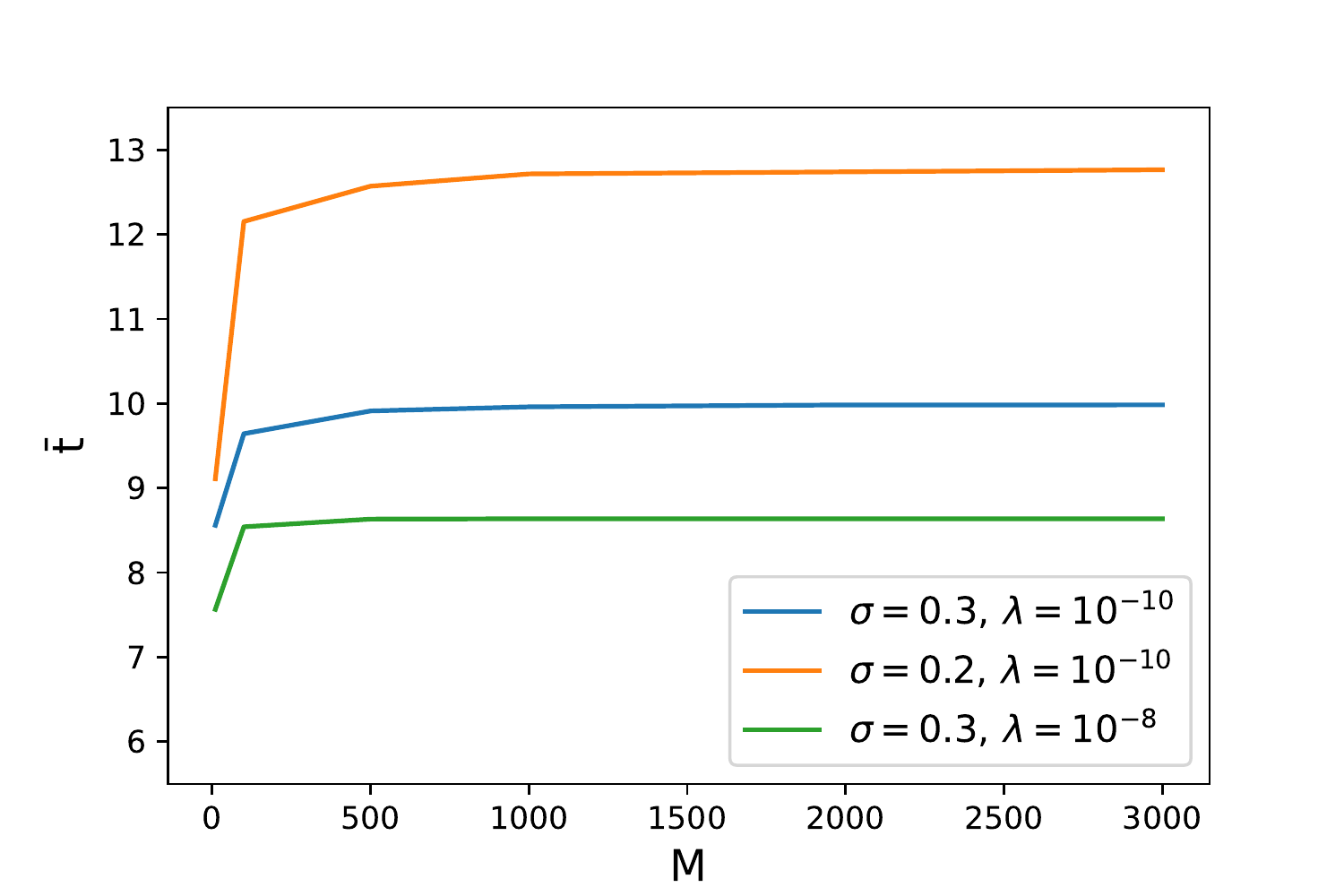} 
\caption{Average test statistics as a function of the number of Nystr\"{o}m centers.}\label{fig:univ_tuning}
\end{subfigure}
\begin{subfigure}{0.49\linewidth}
\centering
\includegraphics[width=\linewidth]{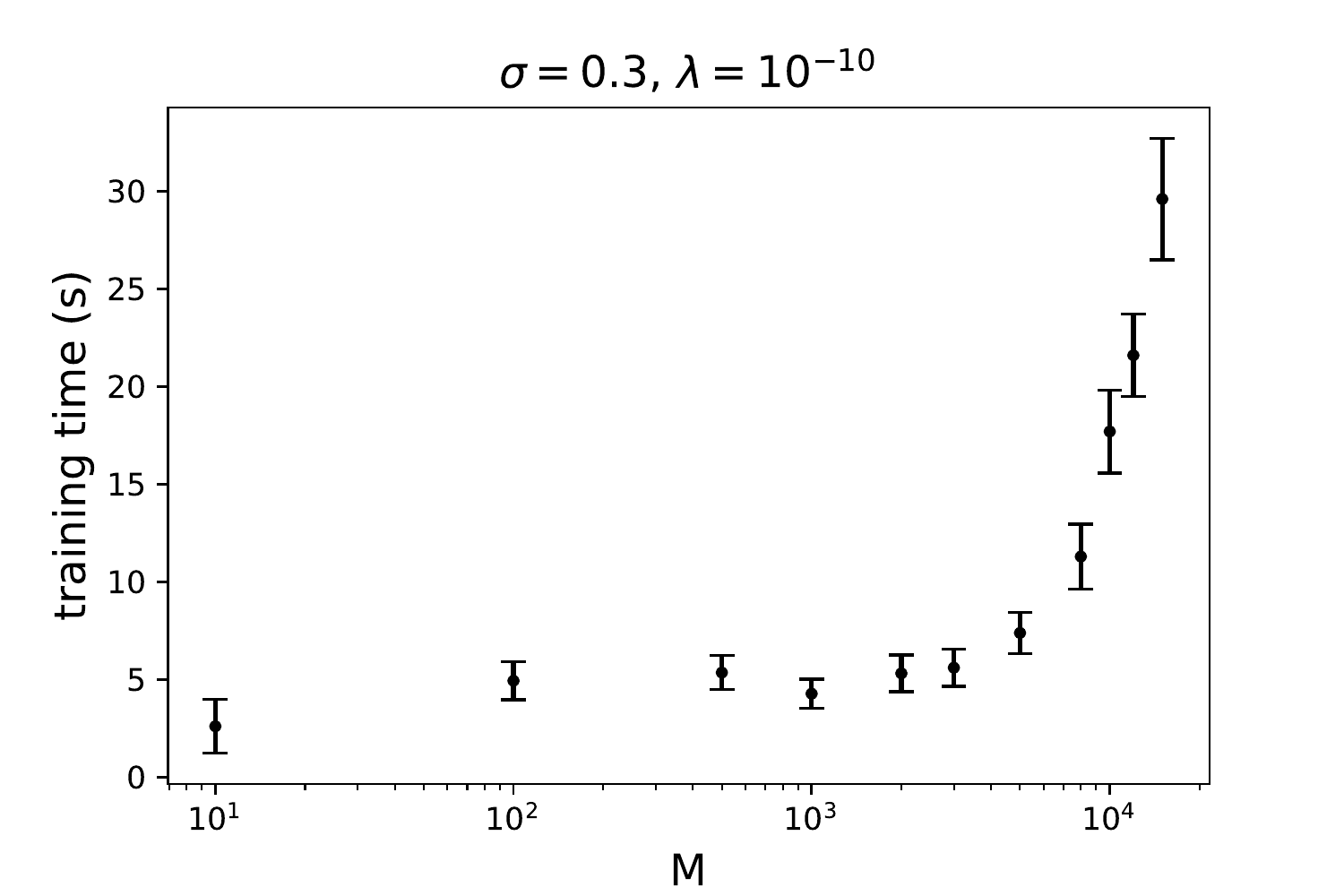} 
\caption{Average training time as a function of the number of Nystr\"{o}m centers.}\label{fig:tr_time_univ}
\end{subfigure}
\caption{}
\label{fig:univ_tuning_all}
\end{figure}

In Figure \ref{fig:univ_tuning} we show that the test statistics averaged over twenty independent runs (with reference data only) reaches a plateau at $M\approx 500\approx \sqrt{\mathN_0}$.  This suggests that the estimated distribution of the test statistics under the null hypothesis does not change if more centers are selected.  On the other hand,  larger values of $M$ might increase the sensitivity of the model to new physics,  at the expense of efficiency in training times and memory.  By looking at the average training time, as reported for instance in Figure \ref{fig:tr_time_univ}, we  fix $M=3000$ .
Figure \ref{fig:univ_tuning} also shows that the value of the test statistics increases for more complex models (smaller $\sigma$ and/or $\lambda$).
Finally,  we take $\lambda=10^{-10}$ because the training would show occasional instabilities at smaller values. 

\begin{figure}[H]
\centering
\includegraphics[scale=0.3]{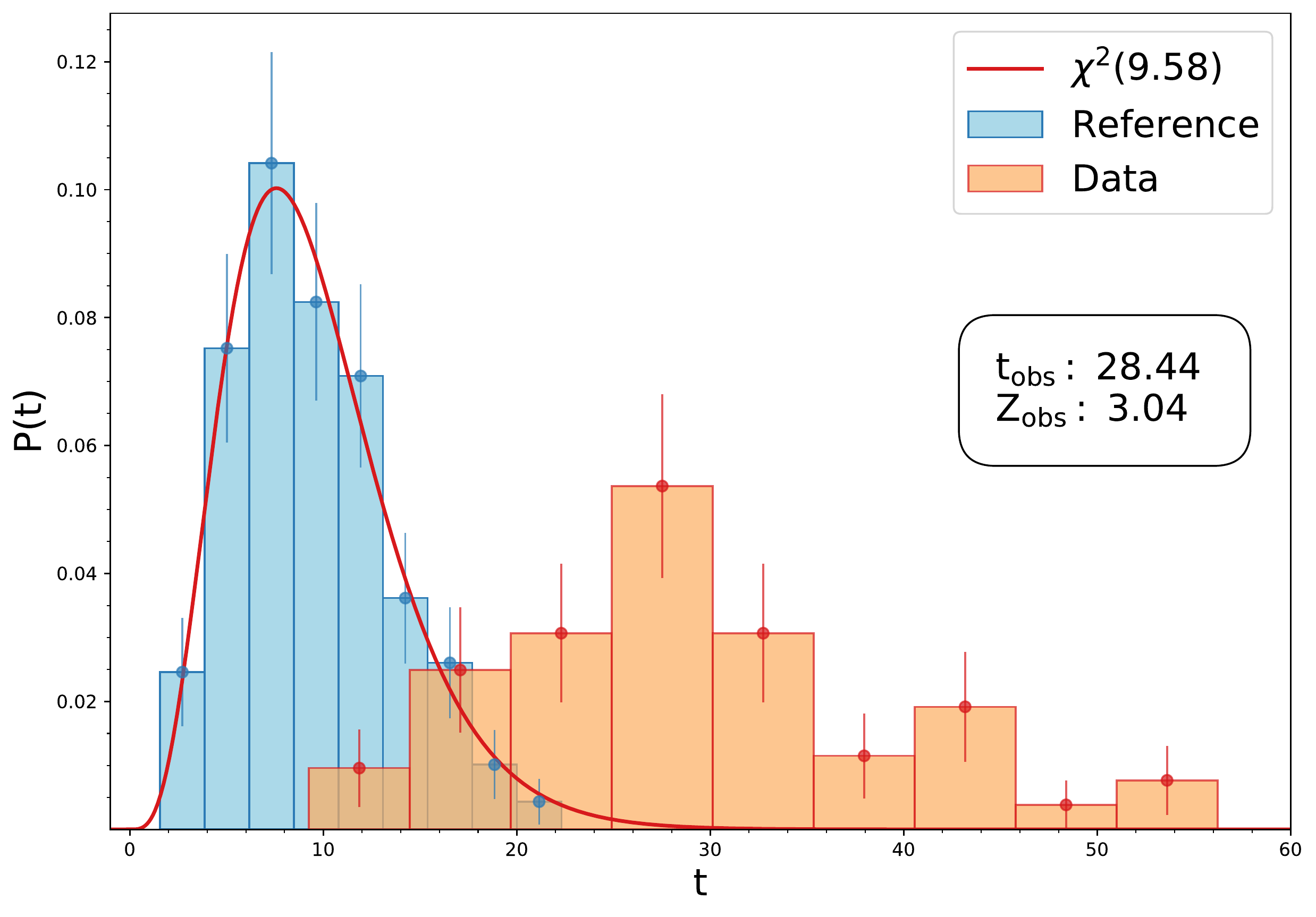}
\caption{Distribution of the test statistics under the null and alternative hypotheses (non-resonant new physics). }
\label{fig:univ_pt}
\end{figure}

\paragraph{Model training and results} 
We first reconstruct the distribution of the test statistics under the null hypothesis $p(t|0)$. 
We train the algorithm on $N_{toys}=300$ toy reference samples $(N(S)=0)$.
In this simple scenario,  we use our complete knowledge of the problem to reconstruct the distribution of the test statistics 
under the alternative hypothesis $p(t|1)$ by performing multiple experiments with $N_{toys}=100$ toy data samples 
with injection of new physics events.  The reconstructed distribution for the non-resonant case is shown in Figure \ref{fig:univ_pt}.  We can see that the test statistics with reference data follows a $\chi^2$ distribution with $9.58$ degrees of freedom (determined with a Kolmogorov-Smironov test).  The median observed significance for the three cases is $Z_{obs} = (2.43,2.82,3.04)$.  The average training time for a single reference toy is $t_{train}\approx 2.11\,s$.  
In this case we can compute the ideal test statistics exactly using the true distributions as follows
\be
t_{id}(S)=-2\left[-N(S)+\sum_{x\in S}\log\left(1+\frac{n(x|S)}{n(x|0)}\right)\right]
\ee
This quantity is then evaluated on a large number (10M) of reference examples to accurately reconstruct $p(t|0)$ and on 300 data samples for each type of signal.  This was done in Ref.~\cite{DAgnolo:2018cun} and the resulting values are $\hat{Z}_{id}=(4.7,  4.1,  4.4)$.  We lose approximately $1.6\sigma$ of sensitivity on average. 

\begin{figure*}
\centering
\begin{subfigure}{.45\linewidth}
\centering
\includegraphics[width=\linewidth]{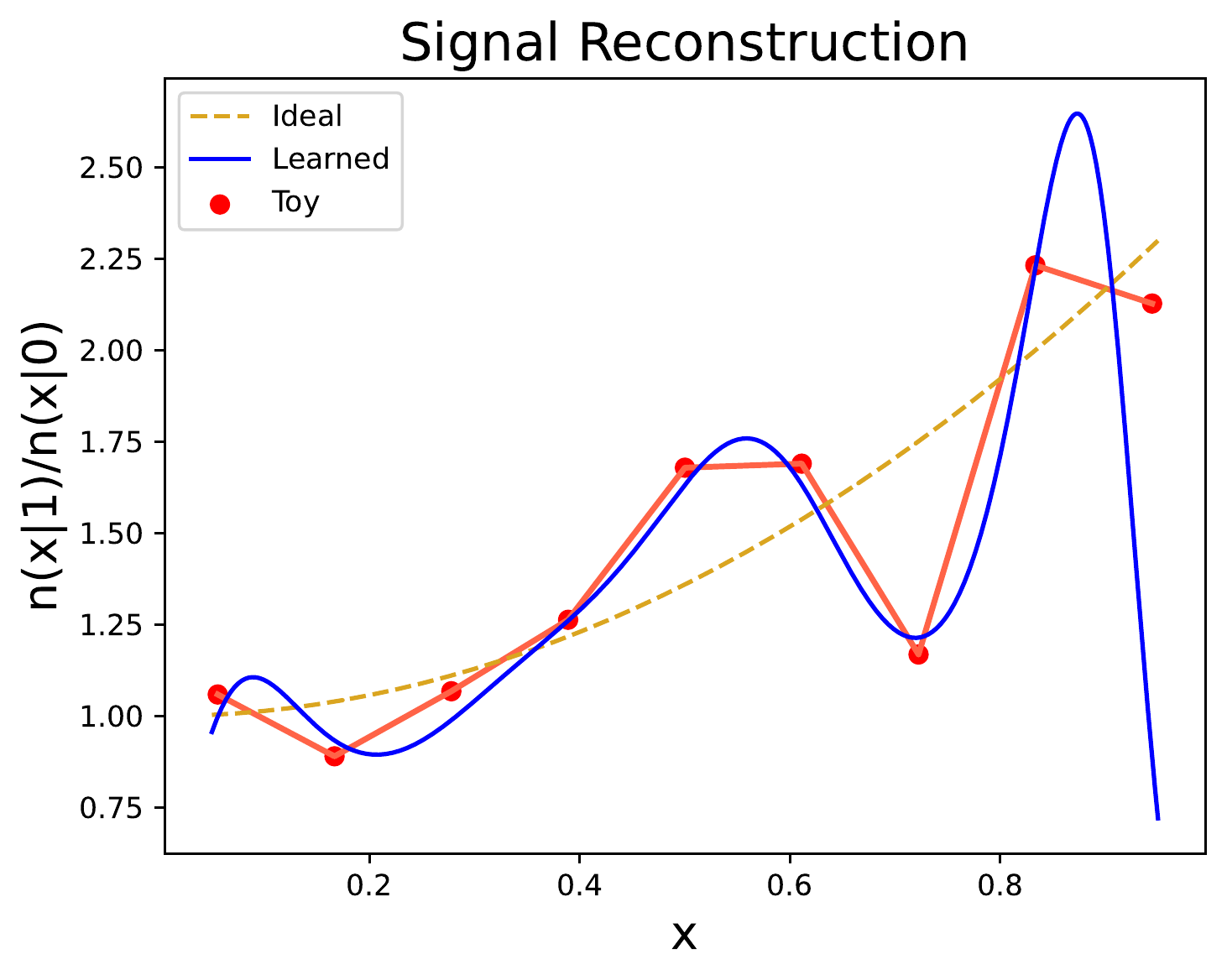}
\end{subfigure}
\centering
\begin{subfigure}{.45\linewidth}
\centering
\includegraphics[width=0.95\linewidth]{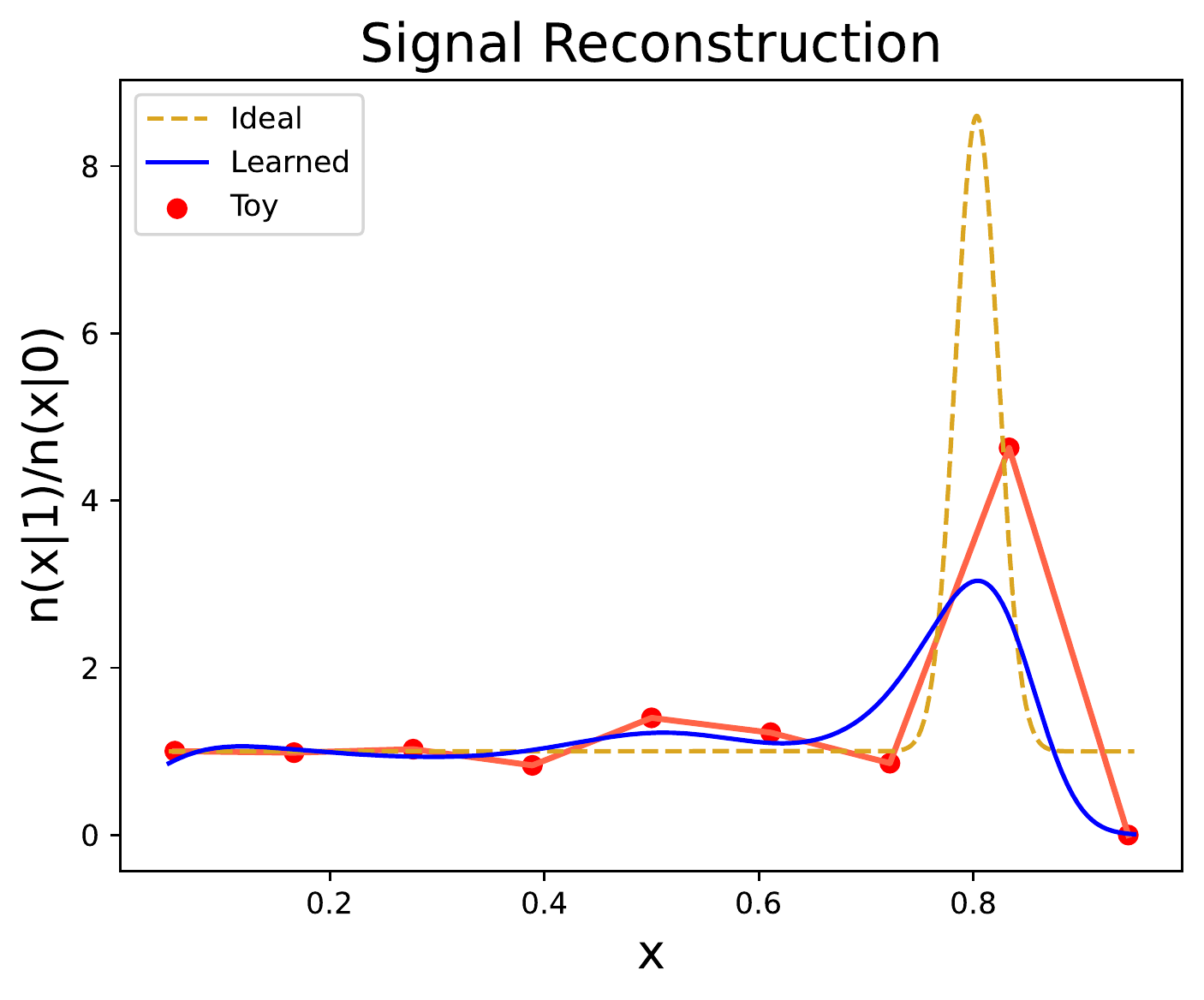}
\end{subfigure}
\caption{Reconstructed density ratios.  Non-resonant signal (left) and Gaussian in the tail (right).}
\label{fig:reco}
\end{figure*}

We can also inspect the learned density ratio to characterize the potential new physics clues.  In this 1D case,  this amounts to simply look at where $\exp(f_{\hat{w}}(x))$ deviates significantly from one.  We show some examples in Figure \ref{fig:reco}.  They are obtained by showing the ideal (exact) likelihood ratio,  the ratio between the (binned) toy and reference samples and the learned functions. 

\begin{figure}[H]
\centering
    \includegraphics[scale=.4]{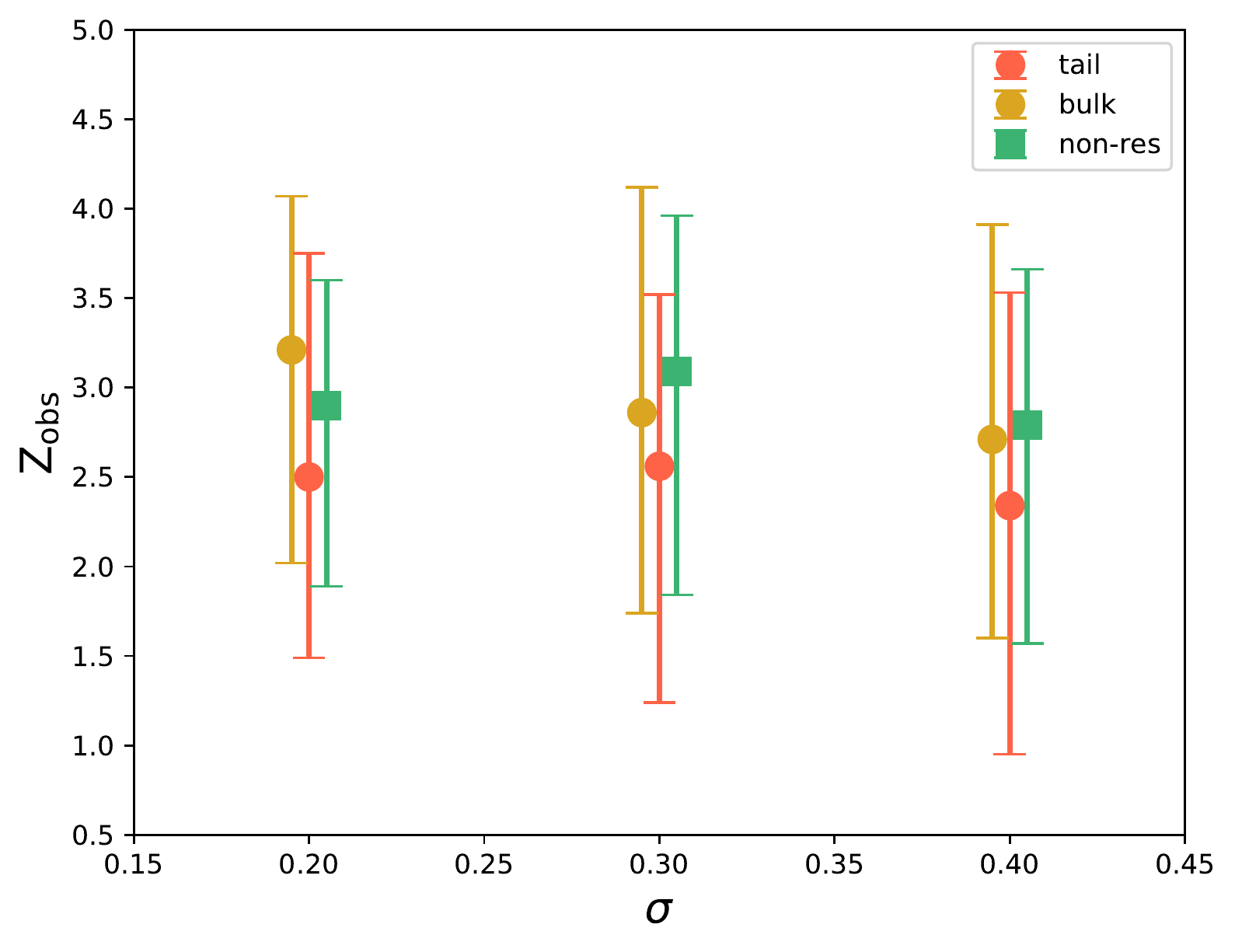}
    \caption{Observed significance  at varying kernel bandwidth.}
    \label{fig:varsigma}
\end{figure}

Finally, Figure \ref{fig:varsigma} shows that the results are stable around the selected bandwidth $\sigma$ across the different types of new physics signals.

\end{document}